\documentclass[12pt]{article}
\usepackage{amsmath, amsfonts, amsthm, amssymb}
\usepackage{natbib}
\usepackage{tikz}
\usepackage{caption, subcaption}
\usepackage{bm} 
\usepackage{graphicx, psfrag, epsf} 
\usepackage{enumerate}
\usepackage{color}
\usepackage{hyperref}
\usepackage{algorithm, algpseudocode}
\usepackage{mathtools}
\usepackage{multirow}
\usepackage{float}
\usepackage{array}
\usepackage[flushleft]{threeparttable}
\usepackage{tabularx}
\usepackage[subpreambles=true]{standalone}
\usepackage{tablefootnote}
\usepackage{standalone}
\makeatletter
\usepackage{xr}
\newtheorem{thm}{Theorem}
\newtheorem{prop}{Proposition}
\newtheorem{cor}{Corollary}
\newtheorem{lemma}{Lemma}
\theoremstyle{definition}
\newtheorem{defn}{Definition}
\def\real{\mathbb R}
\DeclareMathOperator*{\argmin}{arg\,min}

\def\supp{\mathrm{supp}}
\def\st{\mathrm{\quad s.t.\quad}}
\def\T{{\mathcal T}}

\newcommand{\vertiii}[1]{{\left\vert\kern-0.25ex\left\vert\kern-0.25ex\left\vert #1 \right\vert\kern-0.25ex\right\vert\kern-0.25ex\right\vert}}
\algnewcommand\algorithmicinput{\textbf{Input:}}
\algnewcommand\Input{\item[\algorithmicinput]}
\algnewcommand\algorithmicoutput{\textbf{Output:}}
\algnewcommand\Output{\item[\algorithmicoutput]}
\algnewcommand{\algorithmicand}{\textbf{ and }}
\algnewcommand{\algorithmicor}{\textbf{ or }}
\algnewcommand{\OR}{\algorithmicor}
\algnewcommand{\AND}{\algorithmicand}
\newcommand{\StatexIndent}[1][3]{
  \setlength\@tempdima{\algorithmicindent}
  \Statex\hskip\dimexpr#1\@tempdima\relax}
\algdef{S}[IF]{IfNoThen}[1]{\algorithmicif\ #1}
\allowdisplaybreaks 
\def\xVec{\bm x}
\def\yVec{\bm y}

\def\epsilonVec{\bm \varepsilon}
\def\XMat{\bm X}
\def\tildeXMat{\bm{\widetilde X}}
\def\InMat{{\bm I}_n}
\def\IkMat{{\bm I}_k}
\def\IpMat{{\bm I}_p}
\def\ITMat{{\bm I}_{|\T|}}
\def\AMat{\bm A}
\def\ZMat{\bm Z}
\def\UMat{\bm U} 
\def\DMat{\bm D}
\def\VMat{\bm V} 
\def\QMat{\bm Q}
\def\SigmaMat{\boldsymbol \Sigma}
\def\E{{\mathbb E}}
\def\betaVec{\bm \beta}
\def\betaTrueVec{\bm \beta^*}
\def\betaLassoVec{\hat{\bm\beta}^{lasso}_\lambda}
\def\betaOursVec{\hat{\bm\beta}^{ours}_\lambda}
\def\betaLOneAGHVec{\hat{\bm\beta}^{\tt L1-ag-h}_{\lambda, \text{height}}}
\def\betaOLSVec{\hat{\bm\beta}^{\text{OLS}}}
\def\betaAgLS{\hat{\bm\beta^{\mathcal C}}}
\def\tildebetaVec{\tilde{\bm\beta}}
\def\tildebetaTrueVec{\tilde{\bm\beta}^*}
\def\betaOracleVec{{\hat{\bm\beta}^{oracle}_\lambda}}
\def\tildebetaOracleVec{{\check{\bm\beta}^{oracle}_\lambda}}
\def\betaOurVec{\hat{\bm\beta}}
\def\gammaVec{\bm \gamma}
\def\gammaTrueVec{\bm\gamma^*}
\def\gammaOurVec{\hat{\bm\gamma}}
\def\vVec{\bm v}
\def\uVec{\bm u}
\def\betaOneVec{\bm\beta^{(1)}}
\def\betaTwoVec{\bm\beta^{(2)}}
\def\betaThreeVec{\bm\beta^{(3)}}
\def\gammaOneVec{\bm\gamma^{(1)}}
\def\gammaTwoVec{\bm\gamma^{(2)}}
\def\vOneVec{\bm v^{(1)}}
\def\vTwoVec{\bm v^{(2)}}
\def\vThreeVec{\bm v^{(3)}}
\def\uOneVec{\bm u^{(1)}}
\def\uTwoVec{\bm u^{(2)}}
\def\betaOneVecK{\bm {\beta}^{{(1)}k}}
\def\betaOneVecKPlus{\bm {\beta}^{{(1)}k+1}}

\def\betaTwoVecK{\bm {\beta}^{{(2)}k}}
\def\betaTwoVecKPlus{\bm {\beta}^{{(2)}k+1}}

\def\betaThreeVecK{\bm {\beta}^{{(3)}k}}
\def\betaThreeVecKPlus{\bm {\beta}^{{(3)}k+1}}

\def\gammaOneVecK{\bm {\gamma}^{{(1)}k}}
\def\gammaOneVecKPlus{\bm {\gamma}^{{(1)}k+1}}

\def\gammaTwoVecK{\bm {\gamma}^{{(2)}k}}
\def\gammaTwoVecKPlus{\bm {\gamma}^{{(2)}k+1}}

\def\vOneVecK{\bm{v}^{{(1)}k}}
\def\vOneVecKPlus{\bm{v}^{{(1)}k+1}}
\def\vOneVecKMinus{\bm{v}^{{(1)}k-1}}
\def\vTwoVecK{\bm{v}^{{(2)}k}}

\def\vThreeVecK{\bm{v}^{{(3)}k}}

\def\vThreeVecKMinus{\bm{v}^{{(3)}k-1}}
\def\uOneVecK{\bm{u}^{{(1)}k}}
\def\uOneVecKPlus{\bm{u}^{{(1)}k+1}}

\def\uTwoVecK{\bm{u}^{{(2)}k}}

\def\uTwoVecKMinus{\bm{u}^{{(2)}k-1}}
\def\vIVecK{\bm{v}^{{(i)}k}}

\def\vIVecKMinus{\bm{v}^{{(i)}k-1}}
\def\uJVecK{\bm{u}^{{(j)}k}}

\def\uJVecKMinus{\bm{u}^{{(j)}k-1}}
\def\betaIVecK{\bm {\beta}^{{(i)}k}}
\def\betaIVecKPlus{\bm {\beta}^{{(i)}k+1}}

\def\gammaJVecK{\bm {\gamma}^{{(j)}k}}
\def\gammaJVecKPlus{\bm {\gamma}^{{(j)}k+1}}

\def\DeltaVec{\bm \Delta}

\usepackage[margin=1in]{geometry} 
\begin{document}

\def\spacingset#1{\renewcommand{\baselinestretch}%
{#1}\small\normalsize} \spacingset{1}

\title{Rare Feature Selection in High Dimensions}
\author{Xiaohan Yan\thanks{Data Scientist, Microsoft; email: \url{xy257@cornell.edu}} \and Jacob Bien\thanks{Associate Professor, Data Sciences and Operations, Marshall School of Business, University of Southern California; email: \url{jbien@usc.edu}}}
\date{\today}

\maketitle

\begin{abstract}
It is common in modern prediction problems for many predictor variables to be counts of rarely occurring events. This leads to design matrices in which many columns are highly sparse. The challenge posed by such ``rare features'' has received little attention despite its prevalence in diverse areas, ranging from natural language processing (e.g., rare words) to biology (e.g., rare species). We show, both theoretically and empirically, that not explicitly accounting for the rareness of features can greatly reduce the effectiveness of an analysis. We next propose a framework for aggregating rare features into denser features in a flexible manner that creates better predictors of the response. Our strategy leverages side information in the form of a tree that encodes feature similarity.

We apply our method to data from TripAdvisor, in which we predict the numerical rating of a hotel based on the text of the associated review. Our method achieves high accuracy by making effective use of rare words; by contrast, the lasso is unable to identify highly predictive words if they are too rare. A companion {\tt R} package, called {\tt rare}, implements our new estimator, using the alternating direction method of multipliers.
\end{abstract}



\section{Introduction}
\label{sec1}

The assumption of parameter sparsity plays an important simplifying role in high-dimensional statistics.  
However, this paper is focused on sparsity in the data itself, which actually makes estimation more challenging. In many modern prediction problems, the design matrix has many columns that are highly sparse.  This arises when the features record the frequency of events (or the number of times certain properties hold).  While a small number of these events may be common, there is typically a very large number of rare events, which correspond to features that are zero for nearly all observations.  We call these predictors {\em rare features}. Rare features are in fact extremely common in many modern data sets.  For example, consider the task of predicting user behavior based on past website visits: Only a small number of sites are visited by a lot of the users; all other sites are visited by only a small proportion of users.  As another example, consider text mining, in which one makes predictions about documents based on the terms used.  A typical approach is to create a document-term matrix in which each column encodes a term's frequency across documents.  In such domains, it is often the case that the majority of the terms appear very infrequently across the documents; hence the corresponding columns in the document-term matrix are very sparse (e.g., \citealt{forman2003, huang2008, liu2010, wang2010}). In Section \ref{sec6}, we study a text dataset with more than 200 thousand reviews crawled from \url{https://www.tripadvisor.com}. Our goal is to use the adjectives in a review to predict a user's numerical rating of a hotel. As shown in the right panel of Figure \ref{rating_dist_hist}, the distribution of adjective density, defined as the proportion of documents containing an adjective, is extremely right-skewed, with many adjectives occurring very infrequently in the corpus. In fact, we find that more than 95\% of the 7,787 adjectives appear in less than 5\% of the reviews. It is common practice to simply discard rare terms,\footnote{For example, in the {\tt R} text mining library \texttt{tm} \citep{tm2017}, \texttt{removeSparseTerms} is a commonly used function for removing any terms with sparsity level above a certain threshold.} which may mean removing most of the terms (e.g., \citealt{forman2003, huang2008, liu2010, wang2010}). 

Rare features also arise in various scientific fields. For example, microbiome data measure the abundances of a large number of microbial species in a given environment. Researchers use next generation sequencing technologies, clustering these reads into ``operational taxonomic units'' (OTUs), which are roughly thought of as different species of microbe (e.g., \citealt{schloss2009, caporaso2010}). In practice, many OTUs are rare, and researchers often aggregate the OTUs to genus or higher levels (e.g., \citealt{zhang2012, chen2013, xia2013, lin2014, randolph2015, shi2016, cao2017}) or with unsupervised clustering techniques (e.g. \citealt{mcmurdie2013, wang2017b}) to create denser features. However, even after this step, a large portion of these aggregated OTUs are still found to be too sparse and thus are discarded (e.g., \citealt{zhang2012, chen2013, shi2016, wang2017b}). The rationale for this elimination of rare OTUs is that there needs to be enough variation among samples for an OTU to be successfully estimated in a statistical model \citep{ridenhour2017}.

The practice of discarding rare features is wasteful: a rare feature should not be interpreted as an unimportant one since it can be highly predictive of the response.  For instance, using the word ``ghastly'' in a hotel review delivers an obvious negative sentiment, but this adjective appears very infrequently in TripAdvisor reviews.  Discarding an informative word like ``ghastly'' simply because it is rare clearly seems inadvisable.  To throw out over half of one's features is to ignore what may be a huge amount of useful information.

Even if rare features are not explicitly discarded, many existing variable selection methods are unable to select them. The challenge is that with limited examples there is very little information to identify a rare feature as important. Theorem \ref{theorem_ols_limit} shows that even a single rare feature can render ordinary least squares (OLS) inconsistent in the classical limit of infinite sample size and fixed dimension. 

To address the challenge posed by rare features, we propose in this work a method for forming new aggregated features which are less sparse than the original ones and may be more relevant to the prediction task. Consider the following features, which represent the frequency of certain adjectives used in hotel reviews:
\begin{center}
\begin{minipage}{0.6\linewidth}
\begin{itemize}
\item $\XMat_\text{worrying}, \XMat_\text{depressing}, \ldots, \XMat_\text{troubling}$,
\item $\XMat_\text{horrid}, \XMat_\text{hideous}, \ldots, \XMat_\text{awful}$.
\end{itemize}
\end{minipage}
\end{center}
While both sets of adjectives express negative sentiments, the first set (which might be summarized as ``worry'') seems more mild than the second set (which might be summarized as ``horrification''). In predicting the rating of a hotel review, we might find the following two aggregated features more relevant:
\begin{align*}
\tildeXMat_\text{worry} &= \XMat_\text{worrying} + \XMat_\text{depressing}+ \cdots + \XMat_\text{troubling}\\
\tildeXMat_\text{horrification} &= \XMat_\text{horrid} + \XMat_\text{hideous} + \cdots + \XMat_\text{awful}.
\end{align*}
The distinction between ``horrid'' and ``hideous'' might not matter for predicting the hotel rating, whereas the distinction between a ``worry''-related word versus a ``horrification''-related word may be quite relevant. Thus, not only are these aggregated features less rare than the original features, but they may also be more relevant to the prediction task.  A method that selects the aggregated feature $\tildeXMat_\text{horrification}$ thereby can incorporate the information conveyed in the use of ``hideous'' into the prediction task; this same method may be unable to otherwise determine the effect of ``hideous'' by itself since it is too rare.

\begin{figure}
\centering
\includegraphics[width = 0.6\textwidth]{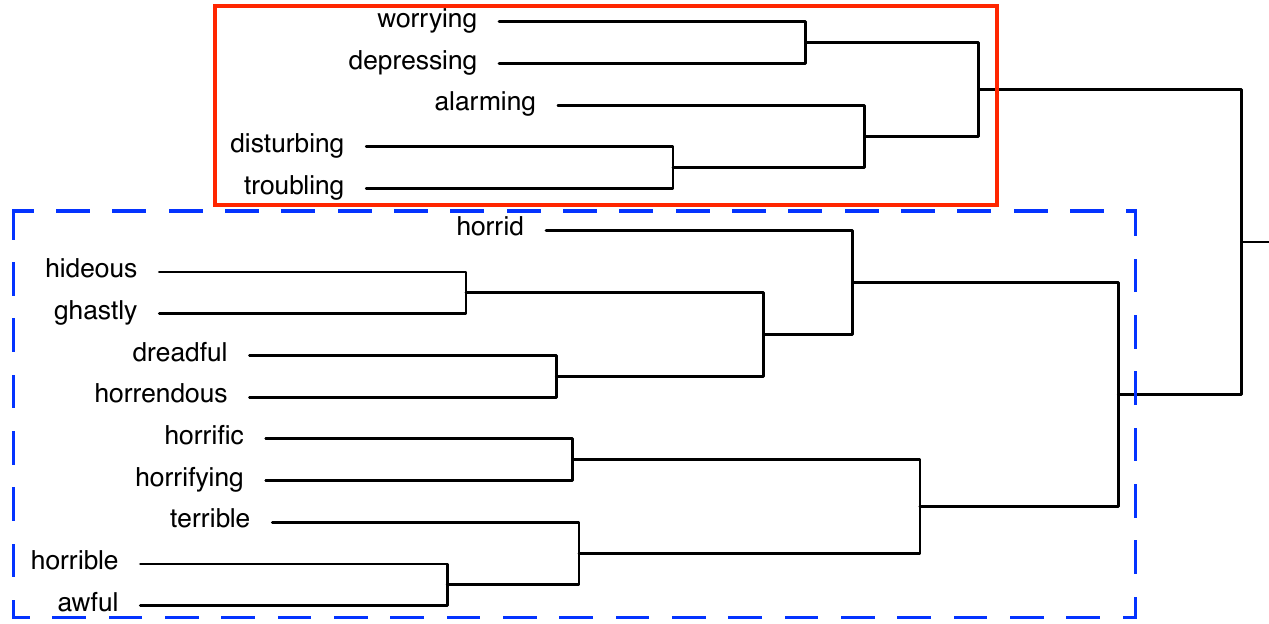}
\caption{A tree that relates adjectives on its leaves}
\label{tree_for_adj}
\end{figure}

Indeed, appropriate aggregation of rare features in certain situations can be key to attaining consistent estimation and support recovery. In Theorem \ref{theorem_lasso_oracle}, we consider a setting where all features are rare and a natural aggregation rule exists among the features. In that setting, we show that the lasso \citep{tibshirani1996} fails to attain high-probability support recovery (for all values of its tuning parameter), whereas an oracle-aggregator does attain this property. Theorem \ref{theorem_lasso_oracle} demonstrates the value of proper aggregation for accurate feature selection when features are rare. This motivates the remainder of the paper, in which we devise a strategy for determining an effective feature aggregation based on data.  Our aggregation procedure makes use of side information about the features, which we find is available in many domains. In particular, we assume that a tree is available that represents the closeness of features.  For example, Figure \ref{tree_for_adj} shows a tree for the previous word example that is generated via hierarchical clustering over \textit{GloVe} \citep{pennington2014} embeddings learned from a different data source. The two contours enclose two subtrees resulting from a cut at their joint node. Aggregating the counts in these subtrees leads to the new features $\tildeXMat_\text{worry}$ and $\tildeXMat_\text{horrification}$ described above. We give more details of constructing such a tree in Section \ref{sec:tree_agg}. 

In Section \ref{sec2}, we motivate our work by providing theoretical results demonstrating the difficulty that OLS and the lasso have with rare features. We further show that correct aggregation of rare features leads to signed support recovery in a setting where the lasso is unable to attain this property.
In Section \ref{sec3}, we introduce a tree-based parametrization strategy that translates the feature aggregation problem to a sparse modeling problem. Our main proposal is an estimator formulated as a solution to a convex optimization problem for which we derive an efficient algorithm.  We draw connections between this method and related approaches and then in Section \ref{sec4}, provide a bound on the prediction error for our method. 
Finally, we demonstrate the empirical merits of the proposed framework through simulation (Section \ref{sec5}) and through the TripAdvisor prediction task (Section \ref{sec6}) described above. In simulation, we examine our method's robustness to misspecified side information.  Quantitative and qualitative comparisons in the TripAdvisor example highlight the advantages of aggregating rare features.

\medskip
\noindent {\bf Notation:} 
Given a design matrix $\XMat\in\real^{n\times p}$, let $\xVec_i\in\real^p$ denote the feature vector of observation $i$ and $\XMat_j\in\real^n$ denote the $j$th feature, where $i=1, \ldots, n$ and $j=1, \ldots, p$. For a vector $\betaVec\in\real^p$, let $\supp(\betaVec)\subseteq\{1,\ldots,p\}$ denote its support (i.e., the set of indices of nonzero elements). Let $\mathbb S_\pm(\betaVec):=\left(sign(\beta_\ell)  \right)_{\ell=1, \ldots, p}$ encode the signed support of the vector $\betaVec$. Let $\T$ be a $p$-leafed tree with root $r$, set of leaves $\mathcal L(\T)=\{1, \ldots, p\}$, and set of nodes $\mathcal V(\T)$ of size $|\T|$. Let $\mathcal T_u$ be the subtree of $\T$ rooted by $u$ for $u\in \mathcal V(\T)$. We follow the commonly-used notions of \textit{child, parent, sibling, descendant}, and \textit{ancestor} to describe the relationships between nodes of a tree. For a matrix $\AMat\in\real^{m\times n}$
and 
subset $B$ of $\{1,\ldots,n\}$, let
 $\AMat_B\in\real^{m\times |B|}$ be the submatrix formed by removing the columns of $\AMat$ not in $B$. Let  $S(\beta_\ell,\lambda)=sign(\beta_\ell)\cdot\max\{|\beta_\ell|-\lambda, 0\}$ be the soft-thresholding operator applied to $\beta_\ell\in\real$.  We let $\bm{e}_j$ denote the vector having a one in the $j$th entry and zero elsewhere.  For sequences $a_n$ and $b_n$, we write $a_n\lesssim b_n$ to mean that $a_n/b_n$ is eventually bounded, and we write $a_n=o(b_n)$ to mean that $a_n/b_n$ converges to zero.


\section{Rare Features and the Promise of Aggregation}
\label{sec2}

\subsection{The Difficulty Posed by Rare Features}
\label{sec:incons-ols-with}

Consider the linear model,
\begin{align}
\yVec=\XMat\betaTrueVec+\epsilonVec,\qquad \epsilonVec\sim N(\bm 0,\sigma^2 \InMat).\label{eq:lm}
\end{align}
where $\yVec=(y_1, \ldots, y_n)^T\in\real^n$ is a response vector, $\XMat\in \real^{n\times p}$ is a design matrix, $\betaTrueVec$ is a $p$-vector of parameters, and $\epsilonVec\in\real^n$ is a vector of independent Gaussian errors having variance $\sigma^2$.  In this paper, we focus on counts data, i.e., $\XMat_{ij}$ records the frequency of an event $j$ in observation $i$. In particular, we will assume throughout that $\XMat$ has non-negative elements.

The lasso \citep{tibshirani1996} is an estimator that performs variable selection, making it well-suited to the $p\gg n$ setting:
\begin{align}
\betaLassoVec\in\argmin_{\betaVec\in{\real^p}}\frac1{2n}\|\yVec-\XMat\betaVec\|_2^2+\lambda\|\betaVec\|_1.\label{eq:lasso}
\end{align}
When $\lambda=0$, this coincides with the OLS estimator, which is uniquely defined when $n>p$ and $\XMat$ is full rank:
$$
\betaOLSVec(n) = (\XMat^T\XMat)^{-1}\XMat^T\yVec.
$$
To better understand the challenge posed by rare features, we begin by considering the effect of a single rare feature on OLS in the classical $p$-fixed, $n\to\infty$ regime. We take the $j$th feature to be a binary vector having $k$ nonzeros, where $k$ is a fixed value  not depending on $n$. As $n$ increases, the proportion of nonzero elements, $k/n$, goes to 0. We show in Theorem \ref{theorem_ols_limit} that $\hat\beta^\text{OLS}_{j}(n)$ does not converge in probability to $\beta^*_j$ with increasing sample size.  This establishes that OLS is not a consistent estimator of $\betaTrueVec$ even in a $p$-fixed asymptotic regime.

\begin{thm}\label{theorem_ols_limit}
Consider the linear model \eqref{eq:lm} with $\XMat\in\real^{n\times p}$ having full column rank.  Further suppose that $\XMat_{j}$ is a binary vector having (a constant) $k$ nonzeros.  It follows that there exists $\eta>0$ for which
$$
\liminf_{n\rightarrow \infty}\mathbb{P}\left(\left|\hat\beta^\text{OLS}_j(n)-\beta^*_j   \right|>\eta  \right) > 0.
$$
\end{thm}
\begin{proof}
The result follows from taking $\liminf_{n\rightarrow \infty}$ of both sides of \eqref{theorem_ols_equation} in Appendix \ref{appen:theorem_ols} and observing that $2\Phi\left(-\eta k^{1/2}/\sigma\right)$ does not depend on $n$.
\end{proof}

The above result highlights the difficulty of estimating the coefficient of a rare feature. This suggests that even when rare features are not explicitly discarded, variable selection methods may fail to ever select them regardless of their strength of association with the response.  
Other researchers have also acknowledged the difficulty posed by rare features in different scenarios. For example, in the context of hypothesis testing for high-dimensional sparse binary regression, \cite{mukherjee2015} shows that when the design matrix is too sparse, any test has no power asymptotically, and signals cannot be detected regardless of their strength.
Since the failure is caused by the sparsity of the features, it is therefore natural to ask if ``densifying the features'' in an appropriate way would fix the problem. As discussed above, aggregating the counts of related events may be a reasonable way to allow a method to make use of the information in rare features.  

\subsection{Aggregating Rare Features Can Help}
\label{sec:aggr-rare-feat}

Given $m$ subsets of $\{1,\ldots,p\}$, we can form $m$ aggregated features by summing within each subset.  We can encode these subsets in a binary matrix $\AMat\in\{0,1\}^{p\times m}$ and form a new design matrix of aggregated features as $\tildeXMat=\XMat \AMat$. The columns of $\tildeXMat$ are also counts, but represent the frequency of $m$ different {\em unions} of the $p$ original events.  For example, if the first subset is $\{1,6,8\}$, the first column of $\AMat$ would be $\bm{e}_1+\bm{e}_6+\bm{e}_8$ and the first aggregated feature would be $\tildeXMat_{1}=\XMat_1+\XMat_6+\XMat_8$, recording the number of times any of the first, sixth, or eighth events occur.  A linear model, $\tildeXMat\tildebetaVec$, based on the aggregated features can be equivalently expressed as a linear model, $\XMat\betaVec$, in terms of the original features as long as $\betaVec$ satisfies a set of linear constraints (ensuring that it is in the column space of $\AMat$):
$$
\tildeXMat \tildebetaVec = (\XMat \AMat)\tildebetaVec=\XMat(\AMat \tildebetaVec)=\XMat\betaVec.
$$
The vector $\betaVec$ lies in the column space of $\AMat$ precisely when it is constant within each of the $m$ subsets.
For example,
\begin{equation}\label{aggregate_equal_coeff}
\small\textit{enforcing }\beta_1 = \beta_6=\beta_8
\quad \Leftrightarrow \quad
\textit{aggregating features: }\XMat_1\beta_1 + \XMat_6\beta_6 + \XMat_8\beta_8 = (\XMat_1+\XMat_6+\XMat_8)\beta_1=\tildeXMat_1\tilde\beta_1.
\end{equation}
In practice, determining how to aggregate features is a challenging problem, and our proposed strategy in Section \ref{sec3} will use side information to guide this aggregation. 

For now, to understand the potential gains achievable by aggregation, we consider an idealized case in which the correct aggregation of features is given to us by an oracle. In the next theorem, we construct a situation in which (a) the lasso on the original rare features is unable to correctly recover the support of $\betaTrueVec$ for any value of the tuning parameter $\lambda$, and (b) an oracle-aggregation of features makes it possible for the lasso to recover the support of $\betaTrueVec$.  For simplicity, we take $\XMat$ as a binary matrix, which corresponds to the case in which every feature has $n/p$ nonzero observations.  We take $\betaTrueVec$ to have $k$ blocks of size $p/k$, with entries that are constant within each block.  The last block is all zeros and the minimal nonzero $|\beta^*_j|$ is restricted to lie within a range that is expands with $n$, $p$ and $k$. The oracle approach delivers to the lasso the $k$ aggregated features that match the structure in $\betaTrueVec$. These aggregated features have $n/k$ nonzeros, and thus are not rare features.  Having peformed the lasso on these aggregated features, we then duplicate the $k$ elements, $p/k$ times per group, to get $\betaOracleVec\in\real^p$.  The lasso with the oracle-aggregator is shown to achieve high-probability signed support recovery whereas the lasso on the original features fails to achieve this property for all values of the tuning parameter $\lambda$.

\begin{thm}\label{theorem_lasso_oracle}
Consider the linear model \eqref{eq:lm}  with binary matrix $\XMat\in\{0,1\}^{n\times p}$, $p\le n$, and $\XMat^T\XMat = (n/p)\IpMat$: every column of $\XMat$ has $n/p$ one's and $\XMat \bm{1}_p = \bm{1}_n$. 
Suppose $\betaTrueVec=\tildebetaTrueVec\otimes \bm{1}_{p/k}$ for $\tildebetaTrueVec=(\tilde\beta^*_1, \ldots, \tilde\beta^*_{k-1}, 0)$. 
Suppose $k<p/(36\log n)$.  Then, the interval $\mathcal I = (\sigma\sqrt{\frac{4k}{n}\log(k^2n)},~\sigma\sqrt{\frac{p}{3n}\log\left(2\tilde c p/k   \right)}]$ with $\tilde c=\frac{1}{3}e^{(\pi/2+2)^{-1}}\sqrt{\frac{1}{4}+\frac{1}{\pi}}$ is nonempty and 
for ${\displaystyle \min_{i=1, \ldots, k-1}\left| \tilde\beta^*_i \right|}\in\mathcal I$, the following two statements hold:
\begin{enumerate}[(a)]
\item The lasso fails to get high-probability signed support recovery:
$$\limsup_{p\rightarrow\infty}\ \sup_{\lambda\ge 0} \mathbb P\left(\mathbb S_\pm(\betaLassoVec)=\mathbb S_\pm(\betaTrueVec)\right)\le\frac{1}{e}.$$
\item The lasso with an oracle-aggregation of features succeeds in recovering the correct signed support for some $\lambda>0$:
$$\lim_{n\rightarrow\infty}\mathbb P\left(\mathbb S_\pm(\betaOracleVec)=\mathbb S_\pm(\betaTrueVec)\right)=1.$$
\end{enumerate}
\end{thm}
\begin{proof}
  See Appendix \ref{appen:theorem_lasso_oracle}.
\end{proof}

Even when the true model does not have a small number of aggregated
features (i.e., $\betaTrueVec$ does not have $k \ll p$ distinct
values), it may still be beneficial to aggregate.  The next result
exhibits a bias-variance tradeoff for feature aggregation.

\begin{thm}[Bias-Variance Tradeoff for Feature
  Aggregated Least Squares]\label{theorem_bias_variance}
  Consider the linear model \eqref{eq:lm}  with $\XMat$ having full
  column rank ($n > p$) and a general vector $\betaTrueVec\in\real^p$.  Let $\mathcal C
  = \{\mathcal C_1,\ldots, \mathcal C_{|\mathcal C|}\}$ be an {\em
    arbitrary} partition of $\{1,\ldots,p\}$.  Let $\betaAgLS\in\real^p$
  be the estimator formed by performing least squares subject to the
  constraint that parameters are constant within the groups defined by
  $\mathcal C$. Then, the following mean squared estimation result
  holds:
  $$
\frac1{n}\E\|\XMat\betaAgLS-\XMat\betaTrueVec\|^2\le\frac1{n}\|\XMat\|_{op}^2 \sum_{\ell=1}^{|\mathcal
    C|}\sum_{j\in\mathcal C_\ell}\left(\beta^*_j - |\mathcal
    C_\ell|^{-1}\sum_{j'\in \mathcal C_\ell}\beta^*_{j'}
  \right)^2+
  \frac{\sigma^2|\mathcal C|}{n}
  $$
\end{thm}
\begin{proof}
  See Appendix \ref{appen:theorem_bias_variance}.
\end{proof}
The bias term is small when there is a small amount of
variability in coefficient values within groups of the partition
$\mathcal C$, i.e. when $\betaTrueVec$ is
{\em approximately} constant within each group.  Even when
$\betaTrueVec$ in truth has a large $k$ (even $k=p$), there may still exist a partition
$\mathcal C$ with $|\mathcal C|\ll k$ for which the bias term is small
and thus $\E\|\XMat\betaAgLS-\XMat\betaTrueVec\|^2\ll
\E\|\XMat\betaOLSVec-\XMat\betaTrueVec\|^2=\sigma^2 p$.


\section{Main Proposal: Tree-Guided Aggregation}
\label{sec3}

In the previous section, we have seen the potential gains achievable through aggregating rare features.  In this section, we propose a tree-guided method for aggregating and selecting rare features.
  We discuss this tree in Section \ref{sec:tree_agg}, introduce a tree-based parametrization strategy in Section \ref{sec:tree_param}, and propose a new estimator in Section \ref{subsubsec.scheme2}.

\subsection{A Tree to Guide Aggregation}
\label{sec:tree_agg}

To form aggregated variables, it is infeasible to consider all possible partitions of the features $\{1,\ldots,p\}$.   Rather, we will consider a tree $\T$ with leaves $1,\ldots,p$ and restrict ourselves to partitions that can be expressed as a collection of branches of $\T$ (see, e.g., Figure \ref{tree_for_adj}).  We sum features within a branch to form our new aggregated features.

We would like to aggregate features that are related, and thus we would like to have $\T$ encode feature similarity information.  Such information about the features comes from prior knowledge and/or data sources external to the current regression problem (i.e., not from $y$ and $X$).  For example, for microbiome data, $\T$ could be the phylogenetic tree encoding evolutionary relationships among the OTUs (e.g., \citealt{matsen2010, tang2016, wang2017a})  or the co-occurrence of OTUs from past data sets.  When features correspond to words, closeness in meaning can be used to form $\T$ (e.g., in Section \ref{sec6}, we perform hierarchical clustering on word embeddings that were learned from an enormous corpus).

In \eqref{aggregate_equal_coeff}, we demonstrated how aggregating a set of features is equivalent to setting these features' coefficients to be equal.  To perform tree-guided aggregation, we therefore associate a coefficient $\beta_j$ with each leaf of $\T$ and ``fuse'' (i.e., set equal to each other) any coefficients within a branch that we wish to aggregate.

\subsection{A Tree-Based Parametrization}
\label{sec:tree_param}

\begin{figure}
\centering
\begin{tikzpicture}[
    scale = 1, transform shape, thick,->,
    every node/.style = {draw, circle, minimum size = 6mm},
    grow = down,  
    level 1/.style = {sibling distance=2cm}, 
    level 2/.style = {sibling distance=0.8cm}, 
    level distance = 1cm
  ]
  \node[fill = white, label = center:$\gamma_8$] (Start) {}
     child { node [fill = white, label = center:$\gamma_6$] {}
     child { node [fill = white, label = center:$\gamma_1$] (1) {}}
     child { node [fill = white, label = center:$\gamma_2$] (2) {}}
     child { node [fill = white, label = center:$\gamma_3$] (3) {}}
   }
     child { node [fill = white, label = center:$\gamma_7$] {}
     child { node [fill = white, label = center:$\gamma_4$] (4) {}}
     child { node [fill = white, label = center:$\gamma_5$] (5) {}}
   };

  \begin{scope}[nodes = {below = 7pt, draw = none}]
      \node [name = X] at (1) {$\beta_1$};
      \node            at (2) {$\beta_2$};
      \node            at (3) {$\beta_3$};
      \node [name = Y] at (4) {$\beta_4$};
      \node            at (5) {$\beta_5$};
  \end{scope}
\end{tikzpicture}
\hspace{1.5cm}
\begin{tikzpicture}[
    scale = 1, transform shape, thick,->,
    every node/.style = {draw, circle, minimum size = 6mm},
    grow = down,  
    level 1/.style = {sibling distance=2cm}, 
    level 2/.style = {sibling distance=0.8cm}, 
    level distance = 1cm
  ]
  \node[fill = white, label = center:$\gamma_8$] (Start) {}
     child { node [fill = white, label = center:$\gamma_6$] {}
     child { node [fill = gray, label = center:$\gamma_1$] (1) {}}
     child { node [fill = gray, label = center:$\gamma_2$] (2) {}}
     child { node [fill = gray, label = center:$\gamma_3$] (3) {}}
   }
     child { node [fill = gray, label = center:$\gamma_7$] {}
     child { node [fill = gray, label = center:$\gamma_4$] (4) {}}
     child { node [fill = gray, label = center:$\gamma_5$] (5) {}}
   };

  \begin{scope}[nodes = {below = 7pt, draw = none}]
      \node [name = X] at (1) {$\beta_1$};
      \node            at (2) {$\beta_2$};
      \node [name = Y] at (3) {$\beta_3$};
      \node [name = U] at (4) {$\beta_4$};
      \node [name = V] at (5) {$\beta_5$};
  \end{scope}
  \draw[densely dashed, rounded corners, thick]
      (X.south west) rectangle (Y.north east);
  \draw[densely dashed, rounded corners, thick]
      (U.south west) rectangle (V.north east);
  
\end{tikzpicture}
\caption{(Left) An example of $\betaVec\in\real^5$ and $\T$ that relates the corresponding five features. By \eqref{betagammarelation}, we have $\beta_i=\gamma_i +\gamma_6 +\gamma_8$ for $i=1,2,3$ and $\beta_j=\gamma_j+\gamma_7+\gamma_8$ for $j=4, 5$. (Right) By zeroing out the $\gamma_i$'s in the gray nodes, we aggregate $\beta$ into two groups indicated by the dashed contours: $\beta_1=\beta_2=\beta_3=\gamma_6+\gamma_8$ and $\beta_4=\beta_5=\gamma_8$. Counts data are aggregated for features sharing the same coefficient: $\XMat\betaVec = (\XMat_1+ \XMat_2 + \XMat_3)\beta_1 + (\XMat_4 + \XMat_5)\beta_4$.  }
\label{twotrees}
\end{figure}
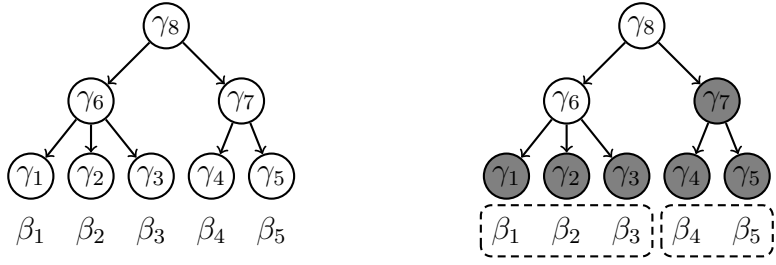

In order to fuse $\beta_j$'s within a branch, we adopt a tree-based parametrization by assigning a parameter $\gamma_u$ to each node $u$ in $\T$ (this includes both leaves and interior nodes). The left panel of Figure \ref{twotrees} gives an example.  Let $ancestor(j)\cup\{j\}$ be the set of nodes in the path from the root of $\T$ to the $j$th feature, which is associated with the $j$th leaf. We express $\beta_j$ as the sum of all the $\gamma_u$'s on the path:
\begin{equation}\label{betagammarelation}
\beta_j=\sum_{u\in ancestor(j)\cup\{j\}} \gamma_u. 
\end{equation}
This can be written more compactly as $\betaVec=\AMat \gammaVec$, where $\AMat \in\{0,1\}^{p\times |\T|}$ is a binary matrix with 
$A_{jk} := 1_{\{u_k\in ancestor(j)\cup\{j\} \}} = 1_{\{j\in descendant(u_k)\cup \{u_k\}\}}$.
The descendants of each node $u$ define a branch $\T_u$, and zeroing out $\gamma_v$'s for all $v\in descendant(u)$ fuses the coefficients in this branch, i.e., $\{\beta_j:j\in\mathcal L(\T_u)\}$.  Thus, $\gamma_{descendant(u)}=0$ is equivalent to aggregating the features $\XMat_j$ with $j\in \mathcal L(\T_u)$ (see the right panel of Figure \ref{twotrees}).  

Another way of viewing this parametrization's merging of branches is by expressing $\XMat\betaVec = \XMat\AMat\gammaVec,$ where $(XA)_{ik}=\sum_{j=1}^p X_{ij}A_{jk}=\sum_{j:j\in descendant(u_k)\cup \{u_k\}} X_{ij}$ aggregates counts over all the descendant features of node $u_k$. By aggregating nearby features, we allow rare features to borrow strength from their neighbors, allowing us to estimate shared coefficient values that would otherwise be too difficult to estimate. In the next section, we describe an optimization problem that uses the $\gammaVec$ parametrization to simultaneously perform feature aggregation and selection.

\subsection{The Optimization Problem}
\label{subsubsec.scheme2}

Our proposed estimator $\hat{\boldsymbol\beta}$ is the solution to the following convex optimization problem:
\begin{equation}\label{ourapproach}
\min_{\substack{\betaVec\in \real^p, \gammaVec \in\real^{|\T|}}}\left\{ \frac{1}{2n} \left\| \yVec - \XMat\betaVec \right\|_2^2 +\lambda\left( \alpha\sum_{\ell = 1}^{|\T|}w_\ell \left| \gamma_\ell\right| +(1-\alpha)\sum_{j=1}^p\tilde w_j \left| \beta_j \right|  \right)  \quad \text{s.t. }\betaVec = \AMat \gammaVec \right\}.
\end{equation}
We apply a weighted $\ell_1$ penalty to induce sparsity in $\gammaOurVec$, which in turn induces fusion of the coefficients in $\betaOurVec$. In the high-dimensional setting, sparsity in feature coefficients is also desirable, so we also apply a weighted $\ell_1 $ penalty on $\betaVec$. The tuning parameter  $\lambda$ controls the overall penalization level while $\alpha$ determines the trade-off between the two types of regularization: fusion and sparsity. In practice, both $\lambda$ and $\alpha$ are determined via cross validation. The choice of weights is left to the user.  Our theoretical results (Section \ref{sec4}) suggest a particular choice of weights, although in our empirical studies (Sections \ref{sec5} and \ref{sec6}) we simply take all weights to be 1 except for the root, which we leave unpenalized ($w_{\text{root}}=0$).  Choosing $w_{\text{root}}=0$ allows one to apply strong shrinkage of all coefficients towards a common value other than zero. 

When $\alpha = 0$, \eqref{ourapproach} reduces to a lasso problem in $\betaVec$; when $\alpha = 1$, \eqref{ourapproach} reduces to a lasso problem in $\gammaVec$. Both extreme cases can be efficiently solved with a lasso solver such as $\mathtt{glmnet}$ \citep{friedman2010}. For $\alpha\in(0,1)$, \eqref{ourapproach} is a generalized lasso problem \citep{tibshirani2011} in $\gammaVec$, and can be solved in principle using preexisting solvers (e.g., \citealt{arnold2014}).   However, better computational performance, in particular in high-dimensional settings, can be attained using an algorithm specially tailored to our problem.  With weights $w_r = 0$ and $\{w_\ell = 1, \tilde w_j =1\}_{\{\ell\neq r, j\in[1, p]\}}$, we write \eqref{ourapproach} as a \textit{global consensus problem} and solve this using alternating direction method of multipliers (ADMM, \cite{boyd2011}). The consensus problem introduces additional copies of $\betaVec$ and $\gammaVec$, which decouples the various parts of the problem, leading to efficient ADMM updates:
\begin{align}\label{consensus}
  &\min_{\substack{\betaOneVec, \betaTwoVec, \betaThreeVec, \betaVec\in \real^p\\\text{and }\gammaOneVec, \gammaTwoVec, \gammaVec\in \real^{|\T|}}} \left\{\frac{1}{2n}\left\|\yVec-\XMat\betaOneVec\right\|_2^2 +\lambda \left( \alpha\sum_{\ell = 1}^{|\T|}w_\ell \left| \gamma_\ell^{(1)}\right| +(1-\alpha)\sum_{j=1}^p\tilde w_j \left| \beta_j^{(2)} \right|  \right)
  \right\}\\
&\text{s.t. } \betaThreeVec = \AMat\gammaTwoVec, \betaVec = \betaOneVec = \betaTwoVec = \betaThreeVec\text{ and }\gammaVec = \gammaOneVec = \gammaTwoVec.\nonumber
\end{align}
In particular, our ADMM approach requires performing a singular value decomposition (SVD) on $\XMat$, an SVD on $(\IpMat:-\AMat)$ (these are reused for all $\lambda$ and $\alpha$), and then applying matrix multiplies and soft-thresholdings until convergence.  See Algorithm \ref{admm} in Appendix \ref{sec:admm} for details. Appendix \ref{appenADMM} provides a derivation of Algorithm \ref{admm} and Appendix \ref{ourapproach_intercept} discusses a slight modification for including an intercept, which is desirable in practice.

\subsection{Connections to Other Work}
\label{related}


Before proceeding, we draw some connections to other work. \citet{wang2017b} introduce a penalized regression method with high-dimensional and compositional covariates that uses a phylogenetic tree. They apply an $\ell_1$ penalty on the sum of coefficients within a subtree, for every possible subtree in the phylogenetic tree. Their penalty is designed to encourage the sum of coefficients within a subtree to be zero, which naturally detects a subcomposition of microbiome features. By contrast, our method applies an $\ell_1$ penalty on our latent variables $\gamma_u$'s, which induces the regression coefficients within subtrees to have equal values. Thus, their penalty encourages the sum of coefficients within a subtree to be zero, whereas ours induces equality. For a simple example, suppose $\beta_1$ and $\beta_2$ form a subtree of the phylogenetic tree.  Their penalty would promote $\beta_1=-\beta_2$ whereas ours would promote $\beta_1=\beta_2$.  The basic assumption of their method is that {\em contrasts} of species abundances within a subtree may be predictive, whereas the basic assumption of our method is that {\em average} species abundance within a subtree may be predictive.  These different structural assumptions will be appropriate in different situations.  In this paper's context of rare features, our assumption is the relevant one: Consider a subtree containing a large number of species.  By asking for equality of coefficients, our method promotes the aggregation of species counts across the subtree, leading to denser features; by contrast, asking for coefficients to sum to zero does not address the problem of feature rarity.

Existing work has also considered the setting in which regression coefficients are thought to be clustered into a small number of groups of equal coefficient value.  For example, \cite{she2010} and \cite{ke2015} do this by penalizing coefficient differences.  Neither method, however, is focused specifically on rare features and thus they do not rely on side information to perform this clustering of coefficients.  In our setting, the side information provided by the tree plays an important role in compensating for the extremely small amount of information available about rare features.

Several other methods assume a relevant undirected graph over the predictors and use a graph-Laplacian or graph-total-variation penalty to promote equality of coefficients that are nearby on the graph \citep{li2010, li2018}. Depending on the setting, this graph may either be pure side information \citep{li2010} or be a covariance graph estimated based on $X$ itself \citep{li2018}.  While the above methods use graph information ``edge-by-edge'', \cite{yu2016} incorporate graphical information ``node-by-node'' to promote joint selection of predictors that are connected on the graph. 

\citet{guinot2017} considers a similar idea of aggregating genomic features with the help of a hierarchical clustering tree; however, in the tree is learned from the design matrix and the prediction task is only used to determine the level of tree cut, whereas our method in effect uses the response to flexibly choose differing aggregation levels across the tree. We consider a strategy similar to theirs, which we call {\tt L1-ag-h} in the empirical comparisons. \citet{kim2012tree} propose a tree-guided group lasso approach in the context of multi-response regression.  In their context, the tree relates the different responses and is used to borrow strength across related prediction tasks. \cite{zhai2018} propose a variance component selection scheme that aggregates OTUs to clusters at higher phylogenetic level, and treats the aggregated taxonomic clusters as multiple random effects in a variance component model. 
Finally, \cite{khabbazian2016} propose a phylogenetic lasso method to study trait evolution from comparative data and detect past changes in the expected mean trait values.


\section{Statistical Theory}
\label{sec4}

In this section, we study the prediction consistency of our method. Since $\T$ encodes feature similarity information, throughout the section we require $\T$ to be a ``full'' tree such that each node is either a leaf or possesses at least two child nodes.
We begin with some definitions.  

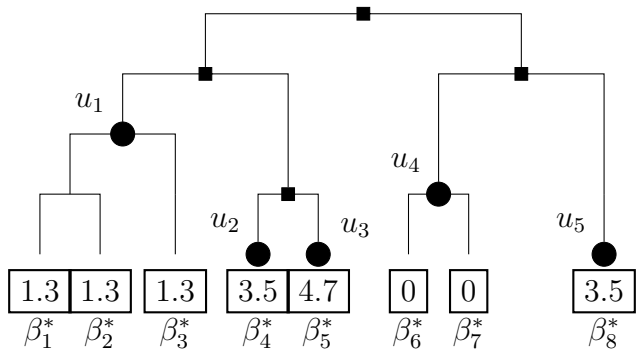
\begin{figure}
\centering
\tikzset{edge from parent fork down/.style={
  edge from parent path={(\tikzparentnode\tikzparentanchor) -- +(0pt,-.0\tikzleveldistance) 
                                                      -| (\tikzchildnode\tikzchildanchor)}}}%
\tikzset{
  solid node/.style={circle,draw,inner sep=3.2,fill=black},
  hollow node/.style={rectangle,draw,inner sep=2.4, fill=black},
}%
\tikzstyle{level 1}=[level distance=0.8cm, sibling distance=4.2cm]%
\tikzstyle{level 2}=[level distance=0.8cm, sibling distance=2.2cm]%
\tikzstyle{level 3}=[level distance=0.8cm, sibling distance=1.4cm]%
\tikzstyle{level 4}=[level distance=0.8cm, sibling distance=0.8cm]%
\begin{tikzpicture}
\node[hollow node](a){}[edge from parent fork down]
	child{
		node [hollow node](b){} 
		child{
			node[solid node, label={[label distance=0.01cm]120: $u_1$}](c){}
			child{
				node[shape=coordinate](d) {}
				child{
					node[shape=coordinate](e) {}
				}
				child{
					node[shape=coordinate](f) {}
				}
			}
			child{ 
				child{
					node[shape=coordinate](g) {}
				}
			}
		}
		child{
			child{
				node[hollow node](h){}
				child{
					node[solid node, label={[label distance=0.01cm]120: $u_2$}](i){}
				}
				child{
					node[solid node, label={[label distance=0.01cm]30: $u_3$}](j){}
				}
			}
		}
	}
	child{
		node[hollow node](k){}
		child{
			child{
				node[solid node, label={[label distance=0.01cm]120: $u_4$}](l){}
				child{
					node[shape=coordinate](m) {}
				}
				child{
					node[shape=coordinate](n) {}
				}
			}
		}
		child{
			child{
				child{
					node[solid node, label={[label distance=0.01cm]120: $u_5$}](o){}
				}
			}
		}
	};
 \begin{scope}[nodes = {below = 6pt, draw = none}]
     \node[name = b1val] at (e) {1.3};
     \node[name = b2val] at (f) {1.3};
     \node[name = b3val] at (g) {1.3};
     \node[name = b4val] at (i) {3.5};
     \node[name = b5val] at (j) {4.7};
     \node[name = b6val] at (m) {0};
     \node[name = b7val] at (n) {0};
     \node[name = b8val] at (o) {3.5};
 \end{scope}
\draw[solid, thick](b1val.south west) rectangle (b1val.north east);
\draw[solid, thick](b2val.south west) rectangle (b2val.north east);
\draw[solid, thick](b3val.south west) rectangle (b3val.north east);
\draw[solid, thick](b4val.south west) rectangle (b4val.north east);
\draw[solid, thick](b5val.south west) rectangle (b5val.north east);
\draw[solid, thick](b6val.south west) rectangle (b6val.north east);
\draw[solid, thick](b7val.south west) rectangle (b7val.north east);
\draw[solid, thick](b8val.south west) rectangle (b8val.north east);

 \begin{scope}[nodes = {below = 6pt, draw = none}]
     \node[name = b1] at (b1val) {$\beta_1^*$};
     \node[name = b2] at (b2val) {$\beta_2^*$};
     \node[name = b3] at (b3val) {$\beta_3^*$};
     \node[name = b4] at (b4val) {$\beta_4^*$};
     \node[name = b5] at (b5val) {$\beta_5^*$};
     \node[name = b6] at (b6val) {$\beta_6^*$};
     \node[name = b7] at (b7val) {$\beta_7^*$};
     \node[name = b8] at (b8val) {$\beta_8^*$};
 \end{scope}

\end{tikzpicture}
\caption{In the above tree, $B^*=\{u_1, u_2, u_3, u_4, y_5\}$ has its nodes labeled with black circles.}
\label{tree_exam_defn}
\end{figure}

\begin{defn}
We say that $B\subseteq\mathcal V(\T)$ is an {\em aggregating set} with respect to $\T$ if $\{\mathcal L(\T_u):u\in B\}$ forms a partition of $\mathcal L(\T)$. 
\end{defn}

The black circles in Figure \ref{tree_exam_defn} form an aggregating set since their branches' leaves are a partition of $\{1,\ldots,8\}$.  We would like to refer to ``the true aggregating set $B^*$ with respect to $\T$'' and, to do so, we must first establish that there exists a unique coarsest aggregating set corresponding to a vector $\betaTrueVec$.

\begin{lemma}\label{lem_aggregating_set}
For any $\betaTrueVec\in\real^p$, there exists a unique coarsest aggregating set $B^*:=B(\betaTrueVec,\T)\subseteq \mathcal V(\T)$ (hereafter ``the aggregating set'') with respect to the tree $\T$ such that 
(a) $\beta^*_j=\beta^*_k$ for $j, k\in \mathcal L(\T_u)$ $\forall u\in B^*$,
(b) $|\beta^*_j-\beta^*_k|>0$ for $j\in \mathcal L(\T_u)$ and $k\in \mathcal L(\T_v)$ for siblings $u, v\in B^*$.
\end{lemma}

The lemma (proved in Appendix \ref{proof_lem_aggregating_set}) defines $B^*$ as the aggregating set such that further merging of siblings would mean that $\beta^*$ is not constant within each subset of the partition. 

\begin{defn}\label{defn_tildebeta}
Given the triplet $(\mathcal T,\betaTrueVec,\XMat)$, we define (a) $ \tildeXMat = \XMat \AMat_{B^*}\in\real^{n\times |B^*|}$ to be the {\em design matrix of aggregated features}, which uses $B^*=B(\betaTrueVec,\T)$ as the aggregating set, and (b) $\tildebetaTrueVec\in \real^{|B^*|}$ to be the coefficient vector using these aggregated features: 
$\betaTrueVec = \AMat_{B^*}\tildebetaTrueVec$.
\end{defn}

We are now ready to provide a bound on the prediction error of our estimator, which is proved in Appendix \ref{proof_maintheorem_alpha2}.


\begin{thm}[Prediction Error Bound]\label{maintheorem_alpha2}
If we take $\lambda \ge 4\sigma\sqrt{\log(2p)/n}$, $\tilde w_j = \left\| \XMat_j \right\|_2/\sqrt{n}$ for $1\le j\le p$ and $w_\ell = \left\| \XMat \AMat_\ell \right\|_2/\sqrt{n}$ for $1\le \ell \le |\T|$, then
$$
\frac{1}{n}\left\| \XMat\betaOurVec - \XMat\betaTrueVec  \right\|_2^2 \le 3\lambda \left( (1-\alpha)\sum_{j\in\mathcal A^*} \tilde w_j |\beta^*_j| + \alpha\sum_{\ell\in B^*}w_\ell |\tilde\beta^*_\ell|  \right)
$$
holds with probability at least $1-2/p$ for any $\alpha\in [0, 1]$.
\end{thm} 

This is a slow rate bound for our method. The standard slow rate bound for the lasso is $\sigma\sqrt{\log p/n}\|\betaTrueVec\|_1$. The next corollary establishes that our method, for any choice of $\alpha$, achieves this rate.

\begin{cor}\label{cor_lasso_rate}
  Suppose $\|\XMat_j\|_2\le \sqrt{n}$ for $1\le j\le p$. Then, taking $\lambda = 4\sigma\sqrt{\log (2p)/n}$ and using the weights in Theorem \ref{maintheorem_alpha2},
  $$
  \frac1{n}\|\XMat (\betaOurVec - \betaTrueVec)\|_2^2\lesssim \sigma\sqrt{\log p/n}\|\betaTrueVec\|_1
  $$
  holds with probablity at least $1-2/p$ for any $\alpha \in [0, 1]$.
\end{cor}
\begin{proof}
See Appendix \ref{sec:proof_of_cor_lasso_rate}.
\end{proof}

The previous corollary establishes that one does not worsen the rate by using our method over the lasso in a generic setting. But is there an advantage to using our method? Our method is designed to do well in circumstances in which $|B^*|$ is small, that is $\betaTrueVec$ is mostly constant with grouping given by the provided tree. The next corollary considers the extreme case in which $\betaTrueVec$ is constant (and nonzero).

\begin{cor}\label{cor_better_lasso_rate}
Under the conditions of Corollary \ref{cor_lasso_rate}, suppose $\beta^*_1 = \ldots = \beta^*_p\neq 0$.  Taking $\alpha\ge\left[1+\|\XMat \bm 1_{p}\|_2/(p\sqrt{n})\right]^{-1}$, 
$$
\frac{1}{n}\left\| \XMat\betaOurVec - \XMat\betaTrueVec  \right\|_2^2\lesssim \sigma \sqrt{\frac{\log p}{n}} \left\|\betaTrueVec \right\|_1 \cdot\frac{\left\|\XMat \bm 1_{p} \right\|_2}{p\sqrt{n}}
$$
holds with probability at least $1-2/p$. Thus, this improves the lasso rate when $\left\|\XMat \bm 1_p \right\|_2 = o(p\sqrt{n})$.
\end{cor}
\begin{proof}
See Appendix \ref{sec:proof_of_cor_better_lasso_rate}.
\end{proof}

For certain sparse designs, the above condition holds. For example, when $n = p$ and $\XMat = \sqrt{n}\InMat$, $\|\XMat \bm 1_p\|_2 = \sqrt{n}\|\bm 1_n\|_2 = n$ which is $o(p\sqrt{n}) = o(n^{3/2})$. The next proposition considers a more general sparse $\XMat$ scenario.

\begin{prop}\label{lemma_generalX}
Suppose each column of $\XMat$ has exactly $r$ nonzero entries chosen at random and independently of all other columns. Suppose all nonzero entries equal $\sqrt{n/r}$ so that $\|\XMat_j\|_2 = \sqrt{n}$ for every column $1\le j\le p$. If 
$
\frac{8\log(n+1)}{3p}<\frac{r}{n}\rightarrow 0,
$
then $\|\XMat \bm 1_p\|_2/(p\sqrt{n})\to0$ in probability.
\end{prop}
\begin{proof}
See Appendix \ref{sec:proof_of_lemma_generalX}.
\end{proof}

Proposition \ref{lemma_generalX} combined with Corollary \ref{cor_better_lasso_rate} demonstrates that our method can improve the prediction error rate over the lasso.  While this corollary focuses on the rather extreme case in which all coefficients are equal, we expect the result to be generalizable to a wider class of settings.  Indeed, the empirical results of the next section suggest that our method can outperform the lasso in many settings.


\section{Simulation Study}
\label{sec5}

We start by forming a tree $\T$ with $p$ leaves.  To do so, we generate $p$ latent points in $\real$ and then apply hierarchical clustering  (using $\mathtt{hclust}$ in \citet{R} with complete linkage).  We would like the tree to partition the leaves into $k$ clusters of varying sizes and at differing heights in the tree (which will correspond to the {\em true aggregation} of the features).  To do so, we first generate $k$ cluster means $\mu_1, \ldots, \mu_k\in\real$ with $\mu_i = 1/i$.  The first $k/2$ means have $3p/(2k)$ associated latent points each, and the remaining $k/2$ means have $p/(2k)$ associated latent points each. The latent points associated with $\mu_i$ are drawn independently from $N\left(\mu_i, \tau^2 \min_{j}(\mu_i - \mu_j)^2\right)$, where $\tau = 0.05$.

By design, there are $k$ interior nodes in $\T$ corresponding to these $k$ groups, which we index by $B^*$.  We form $\AMat$ corresponding to this tree and generate $\betaTrueVec=\AMat_{B^*}\tilde\betaVec^*$.  We zero out $k\cdot s$ elements of $\tilde\betaVec^*\in\real^{k}$ and draw the magnitudes of the remaining elements independently from a Uniform$(1.5, 2.5)$ distribution. We alternate signs of the nonzero coefficients of $\tilde\betaVec^*$.  The design matrix $\XMat\in\real^{n\times p}$ is simulated from a $\text{Poisson}(0.02)$ distribution, and the response $\yVec\in\real^n$ is simulated from \eqref{eq:lm} with $\sigma=\|\XMat\betaTrueVec\|_2/\sqrt{5n}$.  For every method under consideration, we average its performance over 100 repetitions in all the following simulations.

We consider both low-dimensional ($n=500$,  $p=100$, $s=0$) and high-dimensional ($n=100$,  $p=200$, $s\in\{0.2,0.6\}$) scenarios, in each case taking a sequence of $k$ values up to $p/2$. We apply our method with $\T$ and vary the tuning parameters $(\alpha, \lambda)$ along an 8-by-50 grid of values. We take all weights equal to 1 except that of the root node, which we take as zero (leaving it unpenalized). In the low-dimensional case, we compare our method to {\em oracle least squares}, in which we perform least squares on $\XMat\AMat_{B^*}$. Oracle least squares represents the best possible performance of any method that attempts to aggregate features.  We also include least squares on the original design matrix $\XMat$. In the high-dimensional case, we compare our method to the {\em oracle lasso}, in which the true aggregation $\XMat\AMat_{B^*}$ (but not the sparsity in $\tilde\betaTrueVec$) is known, and to the lasso and ridge regression, which are each computed across a grid of 50 values of the tuning parameter.

\begin{figure}
  \centering
  \begin{tikzpicture}[
    grow=right,
    level 1/.style={sibling distance=1.5cm,level distance=5.5cm},
    level 2/.style={sibling distance=1.5cm, level distance=8cm},
    edge from parent/.style={very thick,draw=black,
        shorten >=5pt, shorten <=5pt},
    edge from parent path={(\tikzparentnode.east) -- (\tikzchildnode.west)},
    kant/.style={text width=2cm, text centered, sloped},
    every node/.style={text ragged, inner sep=1mm},
    punkt/.style={rectangle, rounded corners, shade, top color=white,
    bottom color=white, draw=black, very thick },
    punkt2/.style={rectangle, rounded corners, shade, top color=white,
    bottom color=white, draw=gray!40, very thick }
    ]

\node[punkt2, text width=5.5em] {\textit{Are features aggregated?}}
    child {
        node[punkt] [rectangle split, rectangle split, rectangle split parts=2,
         text ragged] {
                  \nodepart{one}
            	{\tt L1}: lasso
                  \nodepart{two}
            	{\tt L1-dense}: {lasso on dense features}
        }
        edge from parent
            node[kant, below, pos=.6] {No}
    }
    child {
        node[punkt2, text width=8em] {\textit{Is the aggregation supervised?}}
        child {
            node [punkt,rectangle split, rectangle split,
            rectangle split parts=2] {
                \nodepart{one}
                {\tt L1-ag-h}: lasso, aggregated by height
                \nodepart{two}
                {\tt L1-ag-d}: {lasso, aggregated by cluster density}
            }
            edge from parent
                node[below, kant,  pos=.6] {No}
        }
        child {
            node [punkt]{
                our method
            }
            edge from parent
                node[kant, above] {Yes}}
            edge from parent{
                node[kant, above] {Yes}}
    };
\end{tikzpicture}
  \caption{A comparison between our method and four other methods}
  \label{six_methods}
\end{figure}
In addition to the above methods, we compare our method to three other approaches, meant to represent variations of how the lasso is typically applied when rare features are present (see Figure \ref{six_methods} for a schematic). The first approach, which we refer to as {\tt L1-dense}, applies the lasso after first discarding any features that are in fewer than $1\%$ of observations.  The second and third approaches apply the lasso with features aggregated according to $\T$ in an unsupervised manner.  The second approach,  {\tt L1-ag-h}, aggregates features that are in the same cluster after cutting the tree at a certain height. In addition to the lasso tuning parameter, the height at which we cut the tree is a second tuning parameter (chosen along an equally-spaced grid of eight values).    The third approach, {\tt L1-ag-d}, performs merges in a bottom-up fashion along the tree until all aggregated features have density above some threshold. This threshold is an additional tuning parameter (chosen along an equally spaced grid of eight values between 0.01 and 1). The lasso tuning parameter in these methods is always chosen along a grid of 50 values.

\begin{figure}
\centering
\includegraphics[width=0.4\textwidth]{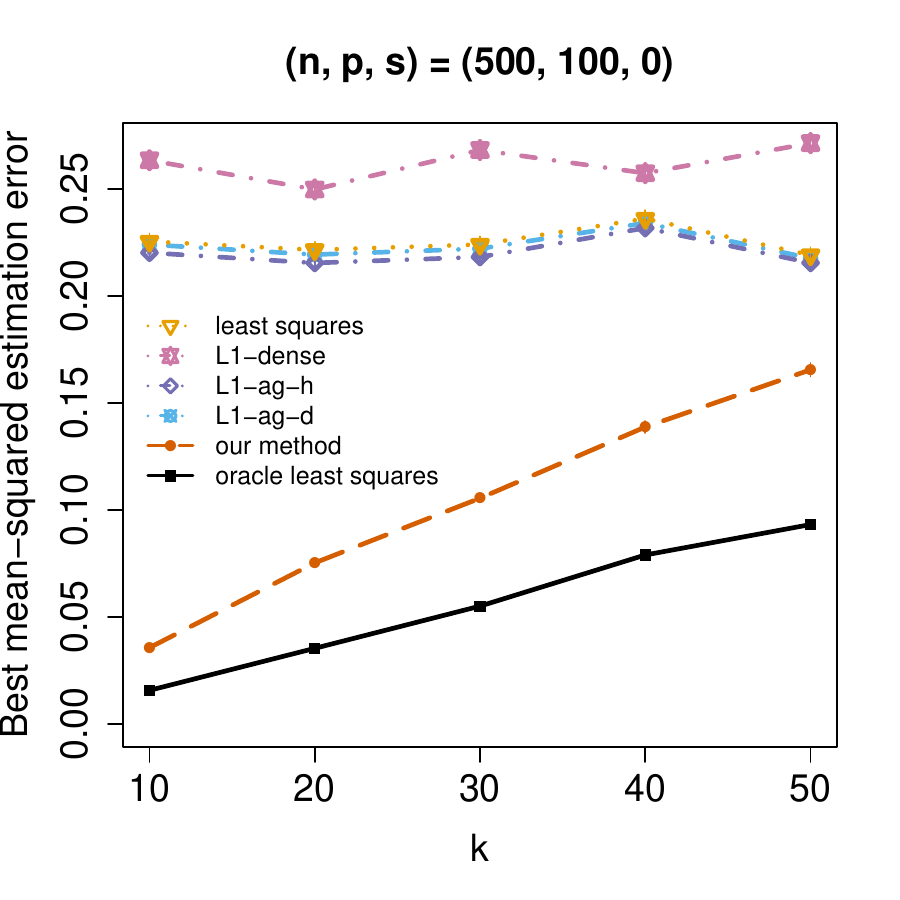}
\includegraphics[width=0.4\textwidth]{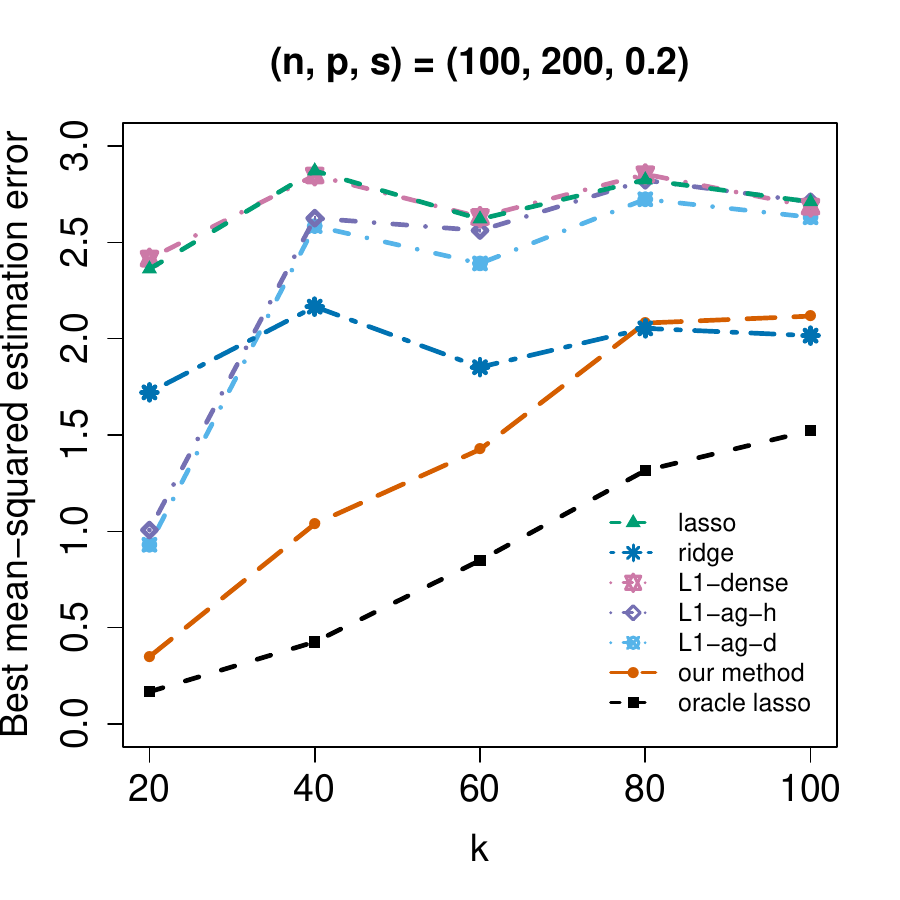}\\
\includegraphics[width=0.4\textwidth]{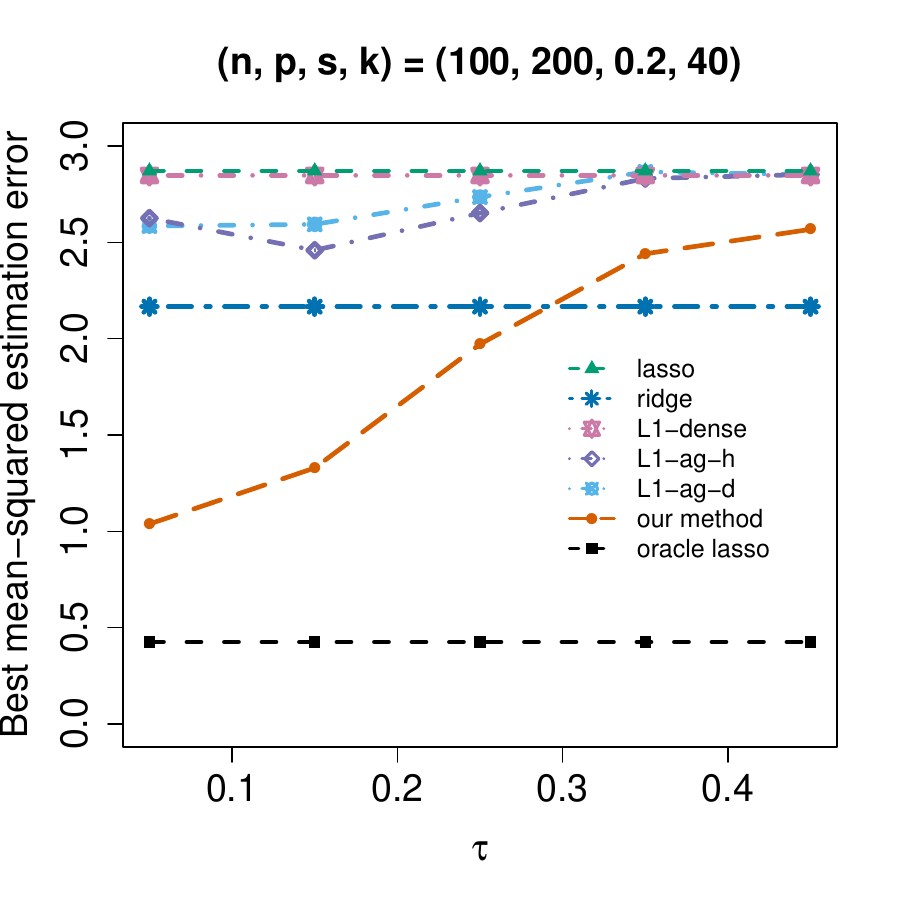}
\includegraphics[width=0.4\textwidth]{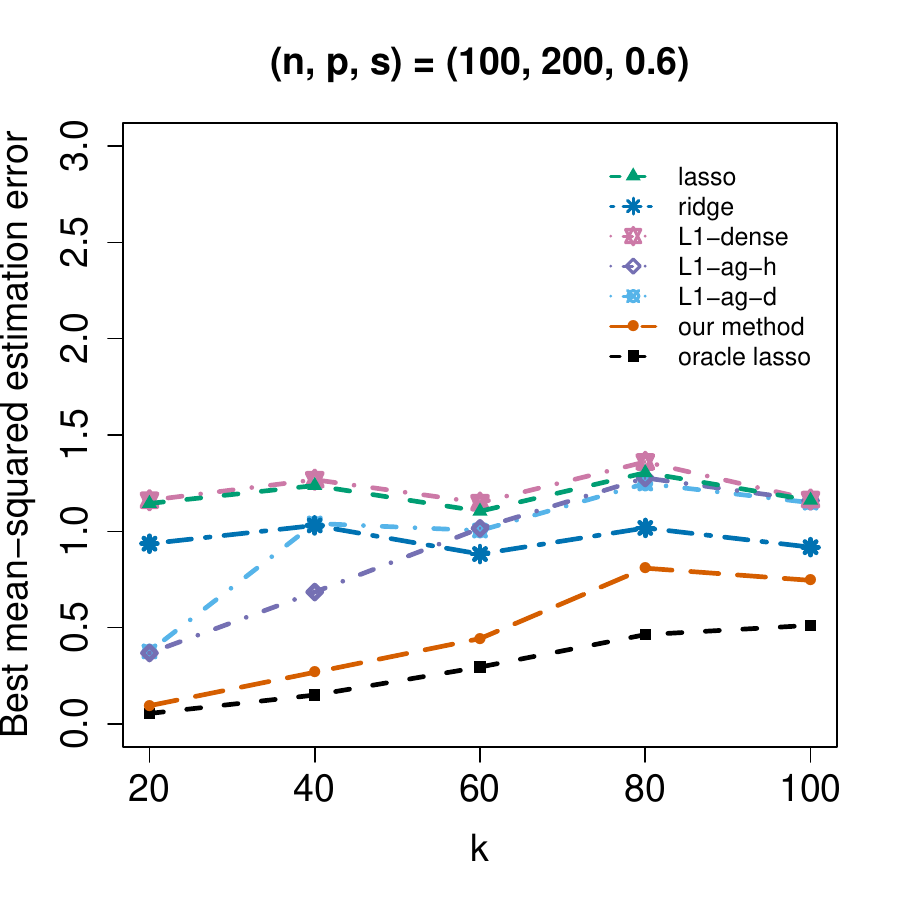}
\caption{Estimation error of all methods under (Top Left) $(n, p, s) = (500, 100, 0)$ versus varying $k\in\{10, 20, 30, 40, 50\}$; (Top Right) $(n, p, s) = (100, 200, 0.2)$ versus varying $k\in\{20, 40, 60, 80, 100\}$; (Bottom Left) $(n, p, s, k) = (100, 200, 0.2, 40)$ versus varying $\tau\in\{0.05, 0.15, 0.25, 0.35, 0.45\}$; (Bottom Right) $(n, p, s) = (100, 200, 0.6)$ versus varying $k\in\{20, 40, 60, 80, 100\}$. }
\label{simu_plot1}
\end{figure}

We measure the {\em best mean-squared estimation error}, i.e., $\min_{\Lambda}\|\betaOurVec(\Lambda) - \betaTrueVec\|_2^2/p$, where ``best'' is with respect to each method's tuning parameter(s) $\Lambda$.  The top two panels of Figure \ref{simu_plot1} show the performance of the methods in the low-dimensional and high-dimensional scenarios, respectively.  Given that our method includes least squares and the lasso as special cases, it is no surprise that our methods have better attainable performance than those methods.  These results indicate that our method performs nearly as well as the oracle when the true number of aggregated features, $k$, is small and degrades (along with the oracle) as this quantity increases. The two other methods that use the tree, {\tt L1-ag-h} and {\tt L1-ag-d}, do less well than our method, but still do better than {\tt L1-dense}, which simply discards rare features. In the $(n, p, s) = (100, 200, 0.2)$ case, {\tt L1-dense} performs almost identically to the lasso, while {\tt L1-ag-h} and {\tt L1-ag-d} degrade to the lasso as $k$ increases. By comparing the right two panels of Figure \ref{simu_plot1}, we notice our method outperforms the ridge at large $k$ when $s$ increases from 0.2 to 0.6, which can be explained by the increased sparsity in $\betaTrueVec$. 

\begin{figure}
\centering
\includegraphics[width=0.4\textwidth]{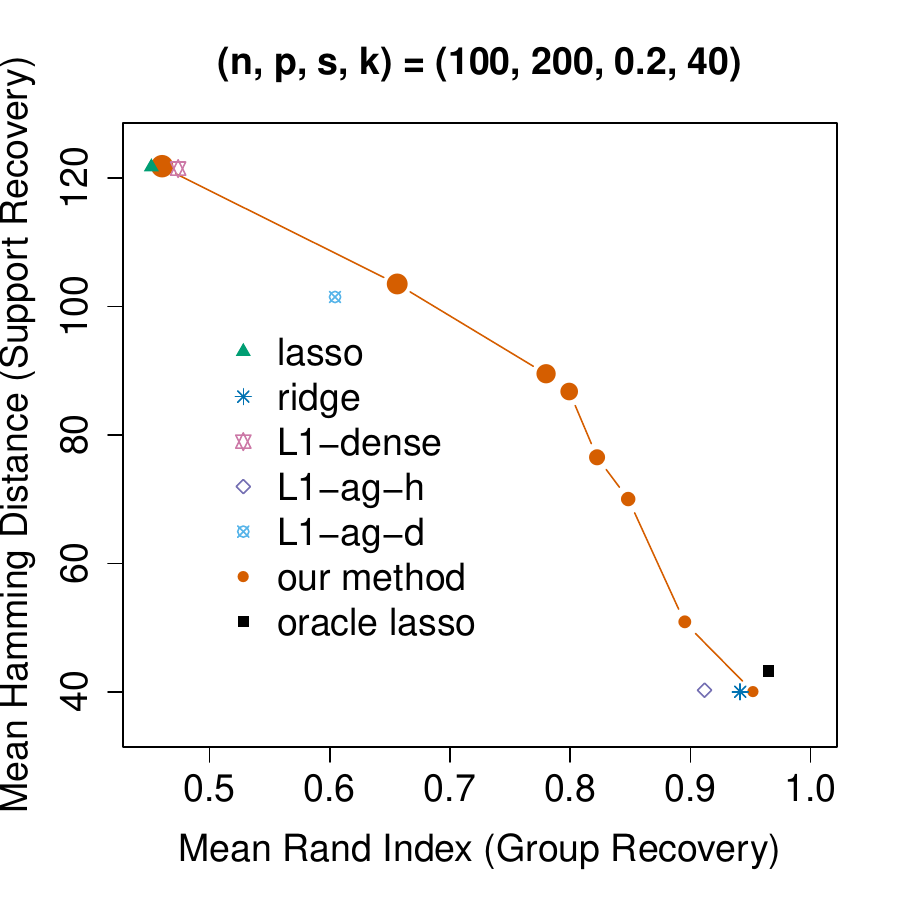}
\includegraphics[width=0.4\textwidth]{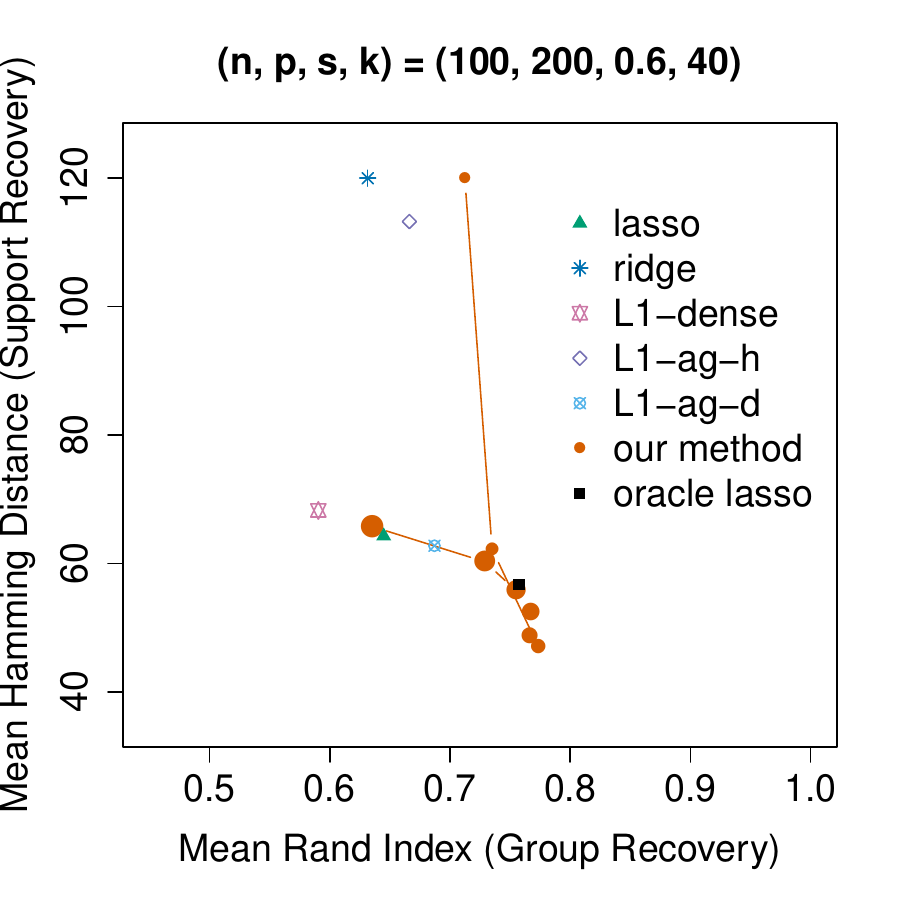}
\caption{Mean Hamming distance versus mean Rand index for various methods' best estimates under (Left) $(n, p, s, k) = (100, 200, 0.2, 40)$ and (Right) $(n, p, s, k) = (100, 200, 0.6, 40)$. For our method, the eight circles correspond to eight different $\alpha$ values, varying from 0 to 1. Decreasing circle size corresponds to increasing $\alpha$ value.}
\label{simu_plot2}
\end{figure}

We also evaluate model performance with respect to {\em group recovery} and {\em support recovery}. Recall that our method is computed over eight $\alpha$ values, between 0 and 1, and fifty $\lambda$ values.  At each $\alpha$, we find the minimizer $\betaOurVec(\hat\lambda)$ for $\|\betaOurVec(\lambda) - \betaTrueVec\|_2^2$ over all $\lambda$ values.  We measure group recovery by computing the Rand index (comparing the grouping of $\betaOurVec$ to that of $\betaTrueVec$), and measure support recovery by computing the Hamming distance (between the supports of $\betaOurVec$ to that of $\betaTrueVec$). The closer the Rand index is to one, the better our method recovers the correct groups. The smaller the Hamming distance is, the better our method recovers the correct support.  For the high-dimensional scenario with $k = 40$, we plot eight pairs (one for each $\alpha$ value) of Rand index and Hamming distance values for $s= 0.2$ and $s= 0.6$ in Figure \ref{simu_plot2}. We also compute the two metrics for the lasso, ridge, {\tt L1-dense}, {\tt L1-ag-h}, {\tt L1-ag-d}, and oracle lasso. In the left panel of Figure \ref{simu_plot2}, which corresponds to the low-sparsity case in $\betaTrueVec$, our method achieves its best performance at the largest $\alpha$ value. As the sparsity level increases, we see from the right panel of Figure \ref{simu_plot2} that the best $\alpha$ shifts towards zero. In both cases, our method outperforms the lasso, ridge, {\tt L1-dense}, {\tt L1-ag-h}, and {\tt L1-ag-d}.

Clearly, the performance of our method, {\tt L1-ag-h} and, {\tt L1-ag-d} will depend on the quality of the tree being used.  In the previous simulations we provided our method with a tree that is perfectly compatible with the true aggregating set.  In practice, the tree used may be only an approximate representation of how features should be aggregated.  We therefore study the sensitivity of our method to misspecification of the tree. We return to the high-dimensional setting above with $k=40$, and we generate a sequence of trees that are increasingly distorted representations of how the data should in fact be aggregated.  

We begin with a true aggregation of the features into $k$ groups as described before. In each repetition of the simulation, we generate a (random) tree $\T$ by performing hierarchical clustering on $p$ random variables generated similarly as before except having increasing $\tau$ value, as a way to control the degradation level of the tree. When $\tau$ is small, the latent variables will be well-separated by group so that the tree will have an aggregating set that matches the true aggregation structure (with high probability).  As $\tau\in \{0.05, 0.15, 0.25, 0.35, 0.45\}$ increases, the between-group variability becomes relatively smaller compared to within-group variability, and thus the information provided by the tree becomes increasingly weak. The bottom left panel of Figure \ref{simu_plot1} shows the degradation of our method as $\tau$ increases when $(n, p, s, k) = (100, 200, 0.2, 40)$. Our method, {\tt L1-ag-h}, and {\tt L1-ag-d} all suffer from a poor-quality tree; the latter two degrade more quickly than ours.


\section{Application to Hotel Reviews}
\label{sec6}

\citet{wang2010} crawled \href{https://www.tripadvisor.com}{TripAdvisor.com} to form a dataset\footnote{Data source: \url{http://times.cs.uiuc.edu/~wang296/Data/}} of 235,793 reviews and ratings of 1,850 hotels by users between February 14, 2009 and March 15, 2009.  While there are several kinds of ratings, we focus on a user's overall rating of the hotel (on a 1 to 5 scale), which we take as our response.  We form a document-term matrix $\XMat$ in which $\XMat_{ij}$ is the number of times the $i$th review uses the $j$th adjective.  

\begin{figure}
\begin{minipage}[b]{0.4\textwidth}
\centering
\begin{tabular}{cc}
Rating & Proportion \\ \hline
1 & 0.066 \\
2 & 0.085 \\
3 & 0.105 \\
4 & 0.308 \\
5 & 0.436 \\ \hline
\end{tabular}
\vspace{0.4in}
\end{minipage}
\begin{minipage}[b]{0.6\textwidth}
\centering
\includegraphics[width=1\textwidth]{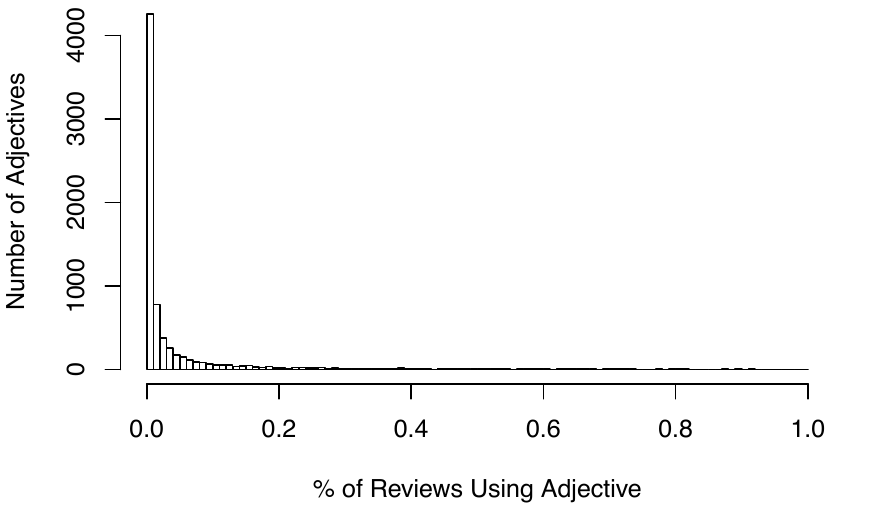}
\end{minipage}
\caption{(Left) distribution of TripAdvisor ratings. (Right) only 414 adjectives appear in more than $1\%$ of reviews;  the histogram gives the distribution of usage-percentages for those adjectives appearing in fewer than $1\%$ of reviews.}
\label{rating_dist_hist}
\end{figure}

We begin by converting words to lower case and keeping only adjectives (as determined by WordNet \citealt{wordnet, wordnet_java, wordnet_r}). After removing reviews with missing ratings, we are left with 209,987 reviews and 7,787 distinct adjectives. The left panel of Figure \ref{rating_dist_hist} shows the distribution of ratings in the data: nearly three quarters of all ratings are above 3 stars. The extremely right-skewed distribution in the right panel of Figure \ref{rating_dist_hist} shows that all but a small number of adjectives are highly rare (e.g., over 90\% of adjectives are used in fewer than $0.5\%$ of reviews).

Rather than discard this large number of rare adjectives, our method aims to make productive use of these by leveraging side information about the relationship between adjectives.  We construct a tree capturing adjective similarity as follows.  We start with word embeddings\footnote{Data source: \url{http://nlp.stanford.edu/data/glove.6B.zip}} in a  100-dimensional space that were pre-trained by \textit{GloVe} \citep{pennington2014} on the Gigaword5 and Wikipedia2014 corpora.  We also obtain a list of adjectives, which the NRC Emotion Lexicon labels as having either positive or negative sentiments \citep{mohammad13}.  We use five nearest neighbors classification within the 100-dimensional space of word embeddings to assign labels to the 5,795 adjectives that have not been labeled in the NRC Emotion Lexicon.  This sentiment separation determines the two main branches of the tree $\T$.  Within each branch, we perform hierarchical clustering of the word embedding vectors.  Figure \ref{dendrogram} depicts such a tree with 2,397 adjectives (as leaves).

\begin{figure}
\centering
\includegraphics[width=1\textwidth]{plots/tripadvisor/dendrogram_1percent.pdf}
\caption{Tree $\T$ over 2,397 adjectives: the left subtree is for adjectives with negative sentiment and the right subtree is for adjectives with positive sentiment.}
\label{dendrogram}
\end{figure}


We compare our method (with weights all equal to 1 except for the root node, which is left unpenalized) to four other approaches described in Figure \ref{six_methods}. For {\tt L1-dense}, we first discard any adjectives that are in fewer than $0.5\%$ of reviews before applying the lasso. Both {\tt L1-ag-h} and {\tt L1-ag-d} have an additional tuning parameter to take care of: for {\tt L1-ag-h} we vary the height at which we cut the tree along an equally-spaced grid of ten values; for {\tt L1-ag-d} we choose the threshold for aggregations' density along an equally spaced grid of ten values between 0.001 and 0.1.


\begin{table}[h]
\spacingset{1} 
\centering
\begin{tabular}{@{\extracolsep{4pt}}rrrr|ccccc@{}}
\empty & \empty &\empty &\empty &\multicolumn{5}{c}{\begin{tabular}{c}Mean Squared Prediction Error \end{tabular}}\\
\cline{5-9} 
prop. & $n$ & $p$ & $n/p$ & {\tt our method} & {\tt L1} & {\tt L1-dense} & {\tt L1-ag-h} & {\tt L1-ag-d} \\
\hline
1\% & 1,700 & 2,397 & 0.71  & \bf 0.870 & 0.894 & 0.895 & 0.882 & 0.971 \\
5\% & 8,499 & 3,962 & 2.15  & \bf 0.783 & 0.790 & 0.805 & 0.785 &0.899 \\
10\% & 16,999 & 4,786 & 3.55 & \bf 0.758 & 0.764 & 0.788 & 0.764 &0.902 \\
20\% & 33,997 & 5,621 & 6.05 & \bf 0.742 & 0.749 & 0.773 & 0.747 & 1.173 \\
40\% & 67,995 &6,472 &10.51 & \bf 0.739 & 0.740 & 0.768 & 0.742 & 1.108 \\
60\% & 101,992 & 6,962 &  14.65 & \bf 0.733 & 0.736 & 0.769 & 0.734 & 1.155 \\
80\% & 135,990 & 7,294 & 18.64 & \bf 0.733 & \bf 0.733 & 0.765 & 0.734 & 0.886 \\
100\% & 169,987 & 7,573 & 22.45 &\bf 0.729 & 0.731 & 0.765 & 0.731 & 0.956 \\
\hline
\end{tabular}
\caption{Performance of five methods on the held-out test set: {\tt L1} is the lasso; {\tt L1-dense} is the lasso on only dense features; {\tt L1-ag-h} is the  lasso with features aggregated based on height; and {\tt L1-ag-d} is the lasso with features aggregated based on density level.}
\label{model_performance}
\end{table}

We hold out 40,000 ratings and reviews as a test set. To observe the performance of these methods over a range of training set sizes, we consider a nested sequence of training sets, ranging from 1\% to 100\% of the reviews not included in the test set. For all methods, we use five-fold cross validation to select tuning parameters and threshold all predicted ratings to be within the interval $[1,5]$.   Table \ref{model_performance} displays the mean squared prediction error (MSPE) on the test set for each method and training set size.

As the size of the training set increases, all methods except for the lasso with aggregation based on density ({\tt L1-ag-d}) achieve lower MSPE. Among the four lasso-related methods, {\tt L1} and {\tt L1-ag-h} outperform the other two. As the training set size $n$ increases, the number of features $p$ also increase but at a relatively slower rate. We notice that when $n/p$ is less than 10.51, our method outperforms the other four lasso-related methods. As $n/p$ increases beyond 10.51, i.e., in the statistically easier regimes, {\tt L1} and {\tt L1-ag-h} attain performance comparable to our method. We conduct paired $t$-tests between squared prediction errors from our method and {\tt L1-ag-h} at every $(n, p)$ pair (i.e., every row of Table \ref{model_performance}). Five out of the eight tests are significant at the 0.005 significance level (See Appendix \ref{sec:pairedttests} Table \ref{tablepairedttests}).

\begin{figure}
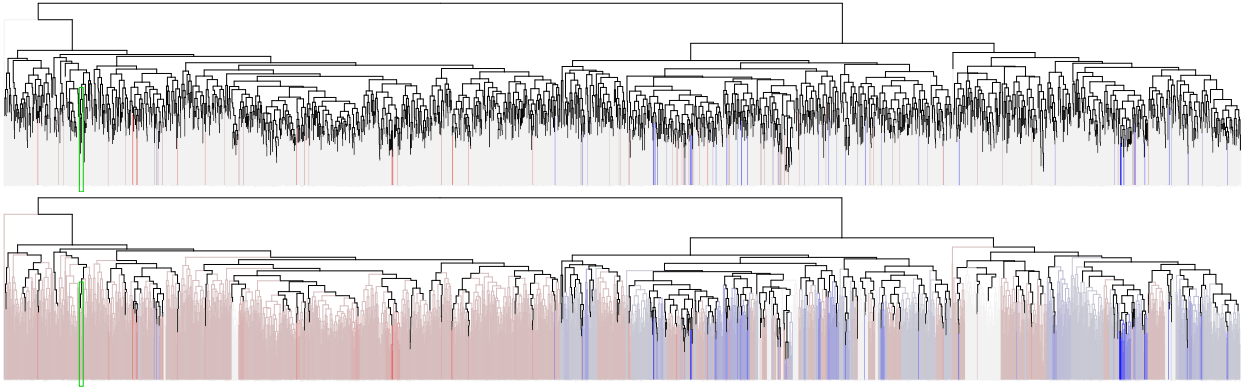

\centering
\includegraphics[width=1\textwidth]{plots/tripadvisor/dendrogram_lasso_1percent.pdf}
\includegraphics[width=1\textwidth]{plots/tripadvisor/dendrogram_l1agh_1percent.pdf}
\includegraphics[width=1\textwidth]{plots/tripadvisor/dendrogram_singlestage_1percent.pdf}
\caption{Trees for $2,397$ adjectives on the leaves with branches colored based on $\betaOurVec$ estimated with the lasso (Top), {\tt L1-ag-h} (Middle) and our method (Bottom), respectively.  {\color{red}Red} branch, {\color{blue}blue} branch and {\color{gray}gray} branch correspond to negative, positive and zero $\hat\beta_j$, respectively. Darker color indicates larger magnitude of $\hat\beta_j$ and lighter color indicates smaller magnitude of $\hat\beta_j$. The horizontal line shown in the middle plot corresponds to the height, chosen from CV, at which {\tt L1-ag-h} cuts the tree and merges the resulting branches.
}
\label{dendrogram_ours_lasso}
\end{figure}

\begin{figure}
\centering
\includegraphics[width=0.8\textwidth]{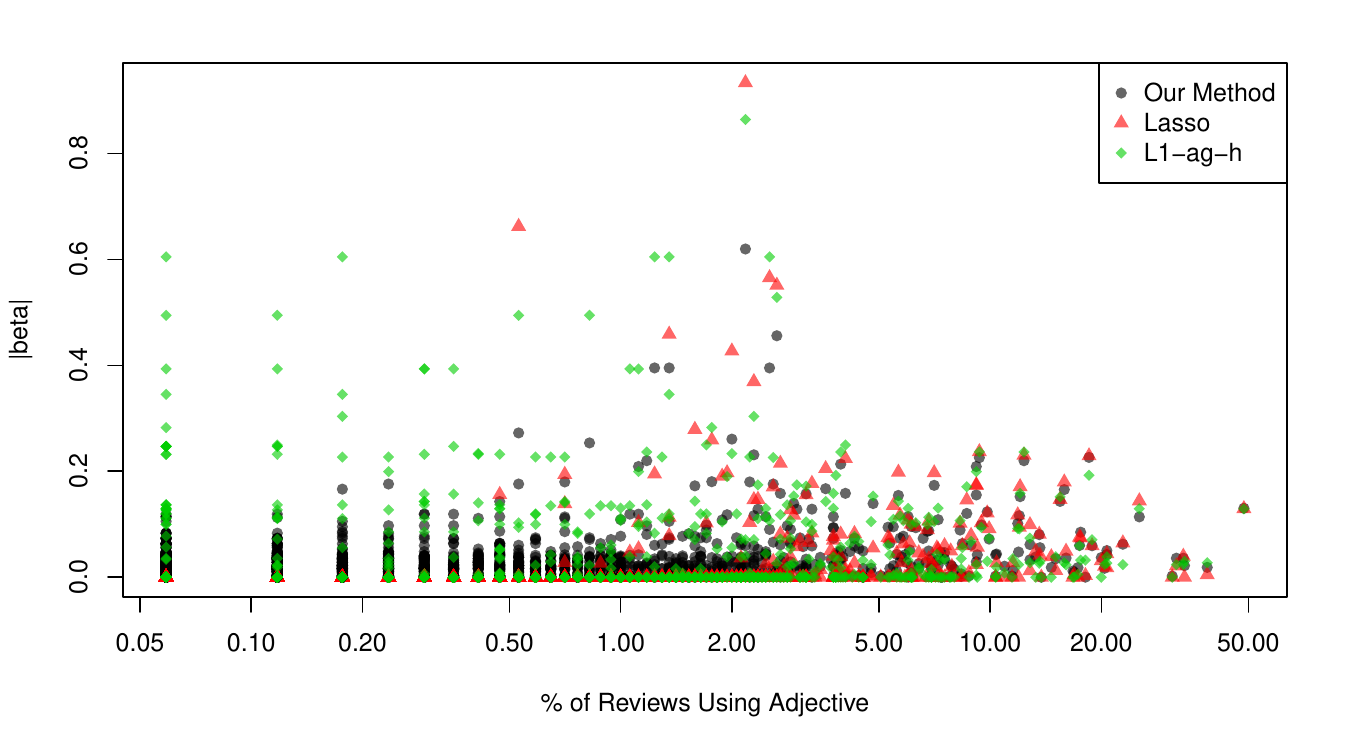}
\caption{Magnitudes of coefficient estimates $\{|\hat\beta_j|\}$ versus feature density (on log scale) for our method (black circles), the lasso (red triangles) and {\tt L1-ag-h} (green diamond).}
\label{coefrarity}
\end{figure}

To better understand the difference between our method, the lasso, and {\tt L1-ag-h}, we color the branches of the tree generated in the $n=1,700$ and $p=2,397$ case (i.e., proportion is 1\%) according to the sign and magnitude of $\betaOurVec$ for the three methods.  The bottom tree in Figure \ref{dendrogram_ours_lasso} corresponds to our method and has many nearby branches sharing the same red/blue color, indicating that the corresponding adjective counts have been merged. By contrast, the top tree in Figure \ref{dendrogram_ours_lasso}, which corresponds to the lasso, shows that the solution is sparser and does not have branches of similar color. The middle tree in Figure \ref{dendrogram_ours_lasso} shows that {\tt L1-ag-h} produces something between the two, merging some adjectives with strong signals while keeping the rest as singletons. The strength and pattern of aggregations vary between our method and {\tt L1-ag-h}. Inspection of the merged branches from our method reveals words of similar meaning and sentiment being aggregated. In Figure \ref{coefrarity}, we plot $\{|\hat\beta_j|\}$ against the percentage of reviews containing an adjective.  We find that our method selects rare words more than the other two methods. The rarest word selected by the lasso is ``filthy'', which appears in 0.47\% of reviews. By contrast, our method selects many words that are far more rare: at the extreme of rarest words, our method selects 797 words that appear in only 0.059\% of reviews; {\tt L1-ag-h} selects 31 out of these 797 rarest words. Our method is able to select more rare words through aggregation: it aggregates 2,244 words into 224 clusters, leaving the remaining 153 words as singletons.  Over 70\% of these singletons are dense words (where, for this discussion, we call a word ``dense'' if it appears in at least 1\% of reviews and ``rare'' otherwise). This is four times higher than the percentage of dense words in the original training data.   Of the 224 aggregated clusters, 42\% are made up entirely of rare words.  After aggregation, over half of these clusters become dense features. As a comparison, {\tt L1-ag-h} aggregates 1,339 words into 506 clusters while keeping 1,058 words as singletons. Only 10\% of these singletons from {\tt L1-ag-h} are dense words. Of the 506 aggregated clusters from {\tt L1-ag-h}, 68\% are made by rare words only, among which only 15\% become dense clusters after aggregation.

Table \ref{qualitative_ex} shows the density and estimated coefficient values for eight words falling in a particular subtree of $\T$.  The words ``heard'' and ``loud'' occur far more commonly than the other six words.  We see that the lasso only selects these two words whereas our method selects all eight words (assigning them negative coefficient values). In contrast, {\tt L1-ag-h}, while also aggregating the two densest words in this branch into a group, does not select the six rare words. Examining the six rare words, it seems quite reasonable that they should receive negative coefficient values.  This suggests to us that the lasso's exclusion of these words has to do with their rareness in the dataset rather than their irrelevance to predicting the hotel rating.  

\begin{table}[h]
\spacingset{1} 
\centering
\begin{tabular}{c|cccccccc}
adjectives & heard & loud & yelled & shouted & screaming & crying & blaring & banging \\ \hline
density \tablefootnote{The term density is computed over the training set.}&0.0300  &  0.0235   & 0.0006 &   0.0006  &  0.0029  &  0.0006  &  0.0006  &  0.0041 \\  \hline       
$\betaLassoVec$& -0.057 &-0.147 & 0&  0&  0&  0&  0&  0\\ \hline
$\betaLOneAGHVec$&-0.174 &-0.174  &0  &0  &0  &0  &0  &0\\ \hline
$\betaOursVec$ &-0.128& -0.128& -0.039& -0.039& -0.039& -0.039& -0.039& -0.039\\ \hline
\end{tabular}
\caption{Term density and estimated coefficient for adjectives in the selected group}
\label{qualitative_ex}
\end{table}

Existing work in sentiment analysis uses side information in other ways to improve predictive performance \citep{thelwall2010}.  For example, \citet{wang2016}, working in an SVM framework, forms a directed graph over words that expresses their ``sentiment strength'' and then requires that the coefficients corresponding to these words honor the ordering that is implied by the directed graph.  For example if word A is stronger (in expressing a certain sentiment) than word B, which in turn is stronger than word C, then their method enforces $\beta_A \ge \beta_B \ge \beta_C$.  Such constraints can in some situations have a regularizing effect that may be of use in rare feature settings: for example, if $\beta_A\approx\beta_C$ and B is a rare word, this constraint would help pin down $\beta_B$'s value.  However, if $\beta_A$ and $\beta_C$ are very different from each other, the constraint may offer little help in reducing the variance of the estimate of $\beta_B$.  Another difference is that the method uses hard constraints that are not controlled by a tuning parameter, so that even when there is strong evidence in the data that a constraint should be violated, this will not be allowed.  By contrast, our method shrinks toward the constant-on-subtrees structure without forcing this to be the case.

\begin{table}[h]
\spacingset{1} 
\centering
\begin{tabular}{l|ccccc}
\empty &\multicolumn{5}{c}{\begin{tabular}{c}Mean Squared Prediction Error \end{tabular}}\\
\cline{2-6} 
tree setting & {\tt our method} & {\tt L1} & {\tt L1-dense} & {\tt L1-ag-h} & {\tt L1-ag-d} \\
\hline
{\it GloVe}-50d& \bf 0.748 & 0.749 & 0.773 & 0.752 & 0.775 \\
{\it GloVe}-100d& \bf 0.742 & 0.749 & 0.773 & 0.747 & 1.173 \\
{\it GloVe}-200d& \bf 0.741 & 0.749 & 0.773 & 0.747 & 1.140 \\
ELMo& \bf 0.669 & 0.676 & 0.732 & 0.685 & 0.783 \\
\hline
\end{tabular}
\caption{Performance of five methods under various tree settings on the held-out test set. Among the tree settings, {\it GloVe}-50d, {\it GloVe}-100d and {\it GloVe}-200d correspond to hierarchical clustering trees generated with {\it GloVe} embeddings of differing dimensions. We collapse over two million ELMo embedding vectors of adjectives into 6,001 clusters using mini-batch K-Means clustering \citep{sculley2010}, then generate a hierarchical tree upon the 6,001 cluster centroids. The performance is based on 33,997 training reviews (i.e., corresponding to the 20\% row of Table \ref{model_performance}).}
\label{model_robustness}
\end{table}

Section \ref{sec5} investigated the effect of using a distorted tree (see the bottom left panel of Figure \ref{simu_plot1}). In the context of words there are multiple choices of trees one could use. Table \ref{model_robustness} shows the effect of applying our method to different types of trees. Here we focus on one situation with 33,997 training reviews. We first consider the effect of changing the dimension of the {\it GloVe} embedding. One might suppose that as the embedding dimension increases the tree becomes more informative. Indeed, one finds that our method and {\tt L1-ag-h} achieve improved performance as this dimension increases.  The last method, {\tt L1-ag-d}, has more variability in its performance as the {\it GloVe} tree varies, making it the poorest performing method. Both {\tt L1} and {\tt L1-dense} do not make use of the tree and thus they are unaffected by changes to the embedding dimension. But a limitation of the methods is that they all are based on the bag-of-words model, and therefore do not take word context into consideration. For example, the models do not differentiate between the use of ``bad'' in ``the hotel is bad'' versus ``the hotel is not that bad''. To bring context into consideration, we leverage deep contextualized word representations from ELMo \citep{peters2018} to generate an auxiliary tree (See Appendix \ref{sec:elmotree} for details). Unlike traditional word embeddings such as \textit{GloVe} that associate one embedding per word, deep word embeddings can capture the meaning of each word based on the surrounding context. From Table \ref{model_robustness} we find test errors improve substantially with the ELMo tree for our method, {\tt L1} and {\tt L1-ag-h}. Among all methods and all tree settings, our method with the ELMo tree performs the best. This suggests our method can better leverage the power of contextualized word embeddings than competing methods.


\section{Conclusion}

In this paper, we focus on the challenge posed by highly sparse data matrices, which have become increasingly common in many areas, including biology and text mining.  While much work has focused on addressing the challenges of high dimensional data, relatively little attention has been given to the challenges of sparsity in the data.  We show, both theoretically and empirically, that not explicitly accounting for the sparsity in the data hurts one's prediction errors and one's ability to perform feature selection.  Our proposed method is able to make productive use of highly sparse features by creating new aggregated features based on side information about the original features.  In contrast to simpler tree-based aggregation strategies that are occasionally used as a pre-processing step in biological applications, our method adaptively learns the feature aggregation in a supervised manner.  In doing so, our methodology not only overcomes the challenges of data sparsity but also produces features that may be of greater relevance to the particular prediction task of interest.

\section*{Acknowledgments}
The authors thank Andy Clark for calling our attention to the challenge of rare features.  This work was supported by NSF CAREER grant, DMS-1653017.




\appendix
\section*{Appendices}
\addcontentsline{toc}{section}{Appendices}
\addcontentsline{toc}{section}{Appendices}


\section{Failure of OLS in the Presence of A Rare Feature}
\label{appen:theorem_ols}

\begin{thm}\label{theorem_ols}
Consider the linear model \eqref{eq:lm} with $\XMat\in\real^{n\times p}$ having full column rank.  Further suppose that $\XMat_{j}$ is a binary vector having $k$ nonzeros.  It follows that
\begin{equation}\label{theorem_ols_equation}
\mathbb{P}\left(\left|\hat\beta^\text{OLS}_j(n)-\beta^*_j   \right|>\eta  \right) \geq 2\Phi\left(-\eta k^{1/2}/\sigma\right)\quad\text{for any }\eta>0,
\end{equation}
where $\Phi(\cdot)$ is the cumulative distribution function of a standard normal variable.
\end{thm}
\begin{proof}
The distribution of the OLS estimator is $\hat\beta^\text{OLS}_j(n)\sim N(\beta^*_j, \sigma^2[(\XMat^T\XMat)^{-1}]_{jj})$. By applying blockwise inversion (see, e.g., \citealt{bernstein2009}), with the $j$th row/column of $\XMat^T\XMat$ in its own ``block'', we get
\begin{align*}
[(\XMat^T\XMat)^{-1}]_{jj} &= [\XMat_j^T\XMat_j - \XMat_j^T\XMat_{-j}(\XMat_{-j}^T\XMat_{-j})^{-1}\XMat_{-j}^T\XMat_j]^{-1}\\
&= [\|\XMat_j\|^2 - \|(\XMat_{-j}^T\XMat_{-j})^{-1/2}\XMat_{-j}^T\XMat_j\|^2]^{-1}\\
&\ge \|\XMat_j\|^{-2}=k^{-1}.
\end{align*}
Thus,
\begin{equation*}
    \mathbb P\left(\left|\hat\beta^\text{OLS}_j(n) -\beta^*_j \right|>\eta \right) =2\Phi\left(-\frac{\eta}{\sigma\sqrt{[(\XMat^T\XMat)^{-1}]_{jj}}}\right)
\ge 2\Phi\left(-\eta k^{1/2}/\sigma\right)
\end{equation*}
where $\Phi(\cdot)$ is the distribution function of a standard normal variable. 
\end{proof}


\section{Proof of Theorem \ref{theorem_lasso_oracle}}
\label{appen:theorem_lasso_oracle}

In the setting of Theorem \ref{theorem_lasso_oracle}, we have $\XMat\betaTrueVec=\tildeXMat \tildebetaTrueVec$,
where $\tildeXMat=\XMat(\IkMat\otimes \bm{1}_{p/k})\in\mathbb R^{n\times
  k}$. The two estimators, the oracle lasso on the aggregated data
($\tildeXMat$) and the lasso on the original data ($\XMat$), are defined below.
\begin{itemize}
\item Oracle lasso estimator $\betaOracleVec=\tildebetaOracleVec \otimes 1_{p/k}$  where $\tildebetaOracleVec$ is the unique solution to 
$$\min_{\tilde \betaVec\in\mathbb R^{k}}\frac{1}{2n}\|\yVec-\tildeXMat \tildebetaVec\|_2^2+\lambda \|\tildebetaVec\|_1.$$
\item Lasso estimator $\betaLassoVec$ is defined in \eqref{eq:lasso}.
\end{itemize}

We begin by establishing that the interval $\mathcal I$ is nonempty,
which is ensured by the constraint $k<p/(36\log n)$.  In particular, the lower bound of the interval is below the upper bound if
$$
12k\log(k^2n)<p\log(2\tilde c p/k) 
$$
Now, $\log(k^2n)\le 3\log n$ and $\log(2\tilde c p/k) \ge 1$ as long as $2\tilde cp/k > e$. Now, $2\tilde c> 0.6$, so if $p/k\ge 5$ then it would suffice to show that
$$
36k\log n<p.
$$
And this constraint does imply that $p/k\ge 5$.  The two parts of the
theorem follow from the following two propositions.

\begin{prop}[Support recovery of oracle lasso]
\label{proporlasso}
Suppose ${\displaystyle \min_{i=1, \ldots, k-1}\left| \tilde\beta^*_i \right|}>\sigma\sqrt{\frac{4k}{n}\log(k^2n)}$. With $\lambda=\sigma\sqrt{\frac{\log(k^2n)}{kn}}$, the oracle lasso recovers the correct signed support successfully:
$$\lim_{n\rightarrow\infty}\mathbb P\left(\mathbb S_\pm(\betaOracleVec)=\mathbb S_\pm(\betaTrueVec)\right)=1.$$
\end{prop}
\begin{proof}
The scaled matrix $\sqrt{\frac{k}{n}}\tildeXMat$ is orthogonal since
\begin{equation*}
\tildeXMat^T\tildeXMat = (\IkMat \otimes \bm{1}_{p/k})^T\XMat^T \XMat (\IkMat \otimes \bm{1}_{p/k}) = \frac{n}{p}\cdot \frac{p}{k}\IkMat = \frac{n}{k}\IkMat.
\end{equation*}
Orthogonality implies that 
\begin{equation}\label{orlasso}
\tildebetaOracleVec = S\left(\left(\sqrt{\frac{k}{n}}\tildeXMat^T\right)\left(\sqrt{\frac{k}{n}}\yVec  \right), \lambda k \right) = S\left(\frac{k}{n}\tildeXMat^T\yVec, \lambda k \right)
\end{equation}
where $\frac{k}{n}\tildeXMat^T\yVec = \frac{k}{n}\tildeXMat^T\tildeXMat \tildebetaTrueVec + \frac{k}{n}\tildeXMat^T\epsilonVec= \tildebetaTrueVec + \frac{k}{n}\tildeXMat^T\epsilonVec\sim N_k(\tildebetaTrueVec, \frac{k\sigma^2}{n}\IkMat)$. By the Chernoff bound for normal variables, for any $t>0$,
$$\mathbb P\left(\left|\frac{k}{n}(\tildeXMat_j)^T\yVec -\tilde\beta^*_j \right| >t \right)\le 2\exp\left(-\frac{t^2}{2k\sigma^2/n}\right)\quad\text{for }j=1, \ldots, k.$$
Choosing $t=\sigma\sqrt{\frac{k\log(k^2n)}{n}}$ and applying a union bound yields
$$\mathbb P\left(\left\| \frac{k}{n}\tildeXMat^T\yVec-\tildebetaTrueVec \right\|_\infty >\sigma\sqrt{\frac{k\log(k^2n)}{n}}  \right)
\le 2k\exp\left(-\frac{\sigma^2k\log(k^2n)/n}{2k\sigma^2/n}  \right)=\frac{2}{\sqrt{n}}. $$
Hence, with probability at least $1-\frac{2}{\sqrt{n}}$, we have $\left\|\frac{k}{n}\tildeXMat^T\yVec-\tildebetaTrueVec \right\|_\infty \le\sigma\sqrt{\frac{k\log(k^2n)}{n}}=\lambda k$, due to our choice of $\lambda=\sigma\sqrt{\frac{\log(k^2n)}{kn}}$. Under $\left\|\frac{k}{n}\tildeXMat^T\yVec-\tildebetaTrueVec \right\|_\infty\le \lambda k$, the following results hold.
\begin{itemize}
\item By $\tilde\beta^*_k=0$ and 
$$\left| \frac{k}{n}(\tildeXMat_k)^T \yVec \right|=\left|\frac{k}{n}(\tildeXMat_k)^T\yVec-\tilde\beta^*_k  \right|\le \left\| \frac{k}{n}\tildeXMat^T\yVec-\tildebetaTrueVec \right\|_\infty \le\lambda k,$$
we have $\check\beta^{oracle}_{\lambda, k}=S\left( \frac{k}{n}(\tildeXMat_k)^T\yVec, \lambda k \right)=0$ and $\check\beta^{oracle}_{\lambda, \ell} = \check\beta^{oracle}_{\lambda, k}=\tilde \beta^*_k=\beta^*_\ell$ for all $\ell>\frac{k-1}{k}n$.
\item For $j=1, \ldots, k-1$, since $\left|\frac{k}{n}(\tildeXMat_j)^T\yVec-\tilde\beta^*_j  \right|\le\lambda k$ and $\left|\tilde \beta^*_j\right|\ge {\displaystyle \min_{i=1, \ldots, k-1}\left| \tilde\beta^*_i \right|} >2\lambda k$, we must have $\frac{k}{n}(\tildeXMat_j)^T\yVec$ and $\tilde\beta^*_j$ share the same sign. Moreover, we either have 
$$\left|\frac{k}{n}(\tildeXMat_j)^T\yVec\right| \ge \left|\tilde\beta^*_j \right|>\lambda k$$
or $\left|\frac{k}{n}(\tildeXMat_j)^T\yVec\right| < \left|\tilde\beta^*_j \right|$ in which case $\left|\frac{k}{n}(\tildeXMat_j)^T\yVec-\tilde\beta^*_j  \right|
=\left|\left|\frac{k}{n}(\tildeXMat_j)^T\yVec\right|-\left|\tilde\beta^*_j \right| \right|
= \left|\tilde\beta^*_j \right|  - \left|\frac{k}{n}(\tildeXMat_j)^T\yVec\right|
\le\lambda k$
and therefore 
$$
\left|\frac{k}{n}(\tildeXMat_j)^T\yVec\right|\ge \left|\tilde\beta^*_j \right|-\lambda k\ge 2\lambda k - \lambda k = \lambda k.
$$
Thus, $\left| \frac{k}{n}(\tildeXMat_j)^T\yVec \right|>\lambda k$ for $j = 1, \ldots, k-1$. By definition of $\betaOracleVec$ and \eqref{orlasso}, for $\frac{j-1}{k}n<\ell \le \frac{j}{k}n$,
$$
\hat\beta^{oracle}_{\lambda, \ell} =\check\beta_{\lambda, j}^{oracle} = S\left(\frac{k}{n}(\tildeXMat_j)^T\yVec, \lambda k  \right) = \frac{k}{n}(\tildeXMat_j)^T\yVec \left(1-\frac{\lambda k}{\left|\frac{k}{n}(\tildeXMat_j)^T\yVec  \right|} \right)
$$
which is of the same sign as $\tilde\beta^*_j$ (and the same sign as $\beta^*_\ell$).
\end{itemize}
In the above two bullet points, we have shown $\mathbb S_\pm(\betaOracleVec)=\mathbb S_\pm(\betaTrueVec)$ holds with probability at least $1-\frac{2}{\sqrt{n}}$. Hence, 
$$
\liminf_{n\rightarrow\infty}\mathbb P\left(\mathbb S_\pm(\betaOracleVec)=\mathbb S_\pm(\betaTrueVec)\right)
\ge \lim_{n\rightarrow\infty}1-\frac{2}{\sqrt{n}}=1.
$$
Since $\limsup_{n\rightarrow\infty}\mathbb P\left(\mathbb S_\pm(\betaOracleVec)=\mathbb S_\pm(\betaTrueVec)\right)=1$, the limit for $\mathbb P\left(\mathbb S_\pm(\betaOracleVec)=\mathbb S_\pm(\betaTrueVec)\right)$ is 1.
\end{proof}

\begin{lemma}\label{lem}
Suppose $\epsilonVec\sim N_n(\bm{0}, \sigma^2\InMat)$ and $\tilde c=\frac{1}{3}e^{(\pi/2+2)^{-1}}\sqrt{\frac{1}{4}+\frac{1}{\pi}}$. Then
$$ \mathbb P\left( \max_{\ell=1, \ldots, p} |\XMat_\ell^T \epsilonVec| \le 2\sigma \sqrt{\frac{n}{3p}\log(2\tilde c p)}  \right)\le \left( 1-\frac{1}{p} \right)^p.$$
\end{lemma}
\begin{proof}
Let $Z$ be a standard Gaussian variable. Theorem 2.1 of \cite{francois2012} provides a lower bound for the Gaussian Q function (i.e., $\mathbb P(Z> z)$). Choosing $\kappa = \frac{3}{2}$ in their Theorem 2.1 yields
$$\mathbb P(Z> z)\ge \underbrace{\left(\frac{1}{3}e^{(\pi/2+2)^{-1}}\sqrt{\frac{1}{4}+\frac{1}{\pi}}\right)}_{\tilde c} e^{-\frac{3z^2}{4}}$$
where $\tilde c=\frac{1}{3}e^{(\pi/2+2)^{-1}}\sqrt{\frac{1}{4}+\frac{1}{\pi}}$ is independent of $z$. Given that $\XMat_\ell$ has $n/p$ one's and $\XMat \bm{1}_p = \bm{1}_n$, we have $\XMat_\ell^T\epsilonVec\stackrel{iid}{\sim}N(0, \frac{n}{p} \sigma^2)$ for $\ell=1, \ldots, p$.
By expressing $\XMat_1^T\epsilonVec=\sqrt{\frac{n}{p}}\sigma Z$, we have for any $\eta>0$,
$$\mathbb P(\XMat_1^T\epsilonVec>\eta)\ge \tilde ce^{-\frac{3\eta^2p}{4\sigma^2n}}\quad\Rightarrow\quad \mathbb P(|\XMat_1^T\epsilonVec|>\eta)\ge 2\tilde ce^{-\frac{3\eta^2p}{4\sigma^2n}}.$$
Moreover, 
$$
\mathbb P\left( \max_{\ell=1, \ldots, p}|\XMat_\ell^T\epsilonVec|\le \eta \right)
=\left( \mathbb P\left( |\XMat_1^T\epsilonVec|\le \eta \right) \right)^p
=\left(1- \mathbb P\left( |\XMat_1^T\epsilonVec|> \eta \right) \right)^p
\le\left(1- 2\tilde ce^{-\frac{3\eta^2p}{4\sigma^2n}} \right)^p.
$$
Plugging in $\eta=2\sigma\sqrt{\frac{n}{3p}\log\left(2\tilde cp  \right)}$ in the above inequality yields
$$
\mathbb P\left( \max_{\ell=1, \ldots, p}|\XMat_\ell^T\epsilonVec|\le 2\sigma\sqrt{\frac{n}{3p}\log\left(2\tilde cp  \right)} \right)
\le \left(1-\frac{1}{p}\right)^p.
$$
\end{proof}

\begin{prop}[Failure of support recovery of lasso]\label{proplasso}
Suppose ${\displaystyle \min_{i=1, \ldots, k-1}\left| \tilde\beta^*_i \right|}\le \sigma\sqrt{\frac{p}{3n}\log\left(2\tilde cp/k   \right)}$ where $\tilde c=\frac{1}{3}e^{(\pi/2+2)^{-1}}\sqrt{\frac{1}{4}+\frac{1}{\pi}}$. The lasso fails to get high-probability signed support recovery: 
$$\limsup_{p\rightarrow\infty}\ \sup_{\lambda\ge 0} \mathbb P\left(\mathbb S_\pm(\betaLassoVec)=\mathbb S_\pm(\betaTrueVec)\right)\le\frac{1}{e}.$$
\end{prop}
\begin{proof}
The lasso solution can be simplified to $\betaLassoVec=S(\frac{p}{n}\XMat^T\yVec, \lambda p)$. Since $\beta^*_\ell\neq 0$ for $\ell\le \frac{k-1}{k}p$ and $\beta^*_\ell=0$ for $\ell>\frac{k-1}{k}p$, the following is a necessary condition for $\betaLassoVec$ to recover the correct signed support:
$$
\exists\ \lambda \text{ s.t. }|\XMat_\ell^T \yVec|>\lambda p \text{ for }\ell\le \frac{k-1}{k}p\text{ and }|\XMat_\ell^T \yVec|\le \lambda p\text{ for }\ell>\frac{k-1}{k}p
$$
$$
\Leftrightarrow
$$
$$
\min_{\ell\le \frac{k-1}{k}p}|\XMat_\ell^T\yVec|>\max_{\ell>\frac{k-1}{k}p}|\XMat_\ell^T\yVec|.
$$
Furthermore, we have 
$$
\XMat_\ell^T \yVec = \XMat_\ell^T \left(\sum_{j=1}^p \XMat_j \beta^*_j + \epsilonVec\right) = \frac{n}{p}\beta^*_\ell + \XMat_\ell^T \epsilonVec.
$$
Define $\bar i := {\displaystyle\argmin_{i=1, \ldots, k-1}\left| \tilde \beta^*_i \right|}$ and $\mathcal A:=\left\{{\displaystyle \max_{\ell>\frac{k-1}{k}p}|\XMat_\ell^T \epsilonVec|}\le 2\sigma\sqrt{\frac{n}{3p}\log(2\tilde c p/k)}  \right\}$. Then
\begin{align*}
&\mathbb P\left(\mathbb S_\pm(\betaLassoVec)=\mathbb S_\pm(\betaTrueVec)\right)\\
\le& \mathbb P\left( \min_{\ell\le \frac{k-1}{k}p}|\XMat_\ell^T \yVec|>\max_{\ell>\frac{k-1}{k}p}|\XMat_\ell^T\yVec| \right)\\
\le& \mathbb P\left( \min_{\frac{\bar i - 1}{k}p <\ell\le \frac{\bar i}{k}p}|\XMat_\ell^T\yVec|>\max_{\ell>\frac{k-1}{k}p}|\XMat_\ell^T\yVec| \right)\\
\le& \mathbb P\left( \frac{n}{p}\left| \tilde\beta^*_{\bar i} \right| + \min_{\frac{\bar i - 1}{k}p <\ell\le \frac{\bar i}{k}p}|\XMat_\ell^T\epsilonVec|>\max_{\ell>\frac{k-1}{k}p}|\XMat_\ell^T\epsilonVec| \right)\\
=& \mathbb P\left( \frac{n}{p}\left| \tilde\beta^*_{\bar i} \right| + \min_{\frac{\bar i - 1}{k}p <\ell\le \frac{\bar i}{k}p}|\XMat_\ell^T\epsilonVec|>\max_{\ell>\frac{k-1}{k}p}|\XMat_\ell^T\epsilonVec| \text{ and } \mathcal A^c\right)\\
&+ \mathbb P\left( \frac{n}{p}\left| \tilde\beta^*_{\bar i} \right| + \min_{\frac{\bar i - 1}{k}p <\ell\le \frac{\bar i}{k}p}|\XMat_\ell^T\epsilonVec|>\max_{\ell>\frac{k-1}{k}p}|\XMat_\ell^T\epsilonVec| \Big | \mathcal A\right)\cdot \mathbb P(\mathcal A)\\
\le& \mathbb P\left( \frac{n}{p}\left| \tilde\beta^*_{\bar i} \right| + \min_{\frac{\bar i - 1}{k}p <\ell\le \frac{\bar i}{k}p}|\XMat_\ell^T\epsilonVec|>2\sigma\sqrt{\frac{n}{3p}\log(2\tilde c p/k)} \right) + \mathbb P\left( \mathcal A \right)\\
=& \left[\mathbb P\left( |\XMat_1^T\epsilonVec|>2\sigma\sqrt{\frac{n}{3p}\log(2\tilde c p/k)} - \frac{n}{p}\left| \tilde\beta^*_{\bar i} \right| \right)\right]^{\frac{p}{k}}+\mathbb P\left( \mathcal A \right)\quad (\text{by }\XMat_\ell^T\epsilonVec \text{ being } i.i.d.)\\
\le& \left[\mathbb P\left( |\XMat_1^T\epsilonVec|>\sigma\sqrt{\frac{n}{3p}\log(2\tilde c p/k)} \right)\right]^{\frac{p}{k}}+\mathbb P\left( \mathcal A \right)\quad\left(\text{by }\left|  \tilde\beta^*_{\bar i}\right|\le \sigma\sqrt{\frac{p}{3n}\log(2\tilde c p/k)}\right)\\
\le& \left[2\exp\left(-\frac{p}{2\sigma^2n}\cdot\frac{\sigma^2n}{3p}\log\left(2\tilde c p/k  \right)   \right)  \right]^{\frac{p}{k}} + \left(1-\frac{k}{p}  \right)^{\frac{p}{k}}\ \left(\text{by Chernoff ineq and Lemma \ref{lem}} \right)\\
\le& 2^{\frac{p}{k}}\exp\left(-\frac{p}{6k}\log\left( 2\tilde c p/k  \right)\right) + \left(1-\frac{k}{p}  \right)^{\frac{p}{k}}\\
=&2^{\frac{p}{k}}\left(\frac{2\tilde c p}{k} \right)^{-\frac{p}{6k}} + \left(1-\frac{k}{p}  \right)^{\frac{p}{k}}\\
=&\left(\frac{\tilde cp}{32k} \right)^{-\frac{p}{6k}} + \left(1-\frac{k}{p}  \right)^{\frac{p}{k}}
\end{align*}
which holds for all $\lambda \ge 0$. In particular, 
$$\sup_{\lambda\ge 0} \mathbb P\left(\mathbb S_\pm(\betaLassoVec)=\mathbb S_\pm(\betaTrueVec)\right)\le \left(\frac{\tilde cp}{32k} \right)^{-\frac{p}{6k}} + \left(1-\frac{k}{p}  \right)^{\frac{p}{k}}.$$
Taking $\limsup$ on both side yields
\begin{align*}
\limsup_{p\rightarrow\infty}\ \sup_{\lambda\ge 0} \mathbb P\left(\mathbb S_\pm(\betaLassoVec)=\mathbb S_\pm(\betaTrueVec)\right)
&\le\lim_{p\rightarrow\infty}\left(\frac{\tilde cp}{32k} \right)^{-\frac{p}{6k}} + \lim_{p\rightarrow\infty}\left(1-\frac{k}{p}  \right)^{\frac{p}{k}}\\
&=0+\frac{1}{e}=\frac{1}{e}.
\end{align*}
\end{proof}


\section{Proof of Theorem \ref{theorem_bias_variance}}
\label{appen:theorem_bias_variance}

Let $\AMat\in\{0,1\}^{p\times |\mathcal C|}$ denote the aggregation matrix
corresponding to the partition $\mathcal C$.  Aggregated least squares
uses the design matrix $\tildeXMat=\XMat\AMat$, where each column is
the sum of features in a group of $\mathcal C$. Left-multiplication
by $\AMat$ maps this back to $p$-dimensional space:
$$
\betaAgLS=\AMat\tildeXMat^{+}\yVec.
$$
Its fitted values are 
\begin{align*}
  \XMat\betaAgLS=\tildeXMat\tildeXMat^{+}\yVec=\tildeXMat\tildeXMat^{+}(\XMat\betaTrueVec+\epsilonVec).
\end{align*}
Now, $\tildeXMat$ is full-rank and
$$
\XMat\betaAgLS\sim N_p(\tildeXMat\tildeXMat^{+}\XMat\betaTrueVec, \sigma^2 \tildeXMat[\tildeXMat^T \tildeXMat]^{-1}\tildeXMat^T).
$$
Now,
\begin{align*}
 \E[\XMat\betaAgLS]&= \tildeXMat\tildeXMat^{+}\XMat\betaTrueVec\\
&=\tildeXMat\tildeXMat^{+}\XMat\AMat\AMat^+\betaTrueVec+\tildeXMat\tildeXMat^{+}\XMat({\bf
    I}_p-\AMat\AMat^+)\betaTrueVec\\
&=\XMat\AMat\AMat^+\betaTrueVec+\tildeXMat\tildeXMat^{+}\XMat({\bf
    I}_p-\AMat\AMat^+)\betaTrueVec
\end{align*}
since $\XMat\AMat\AMat^+\betaTrueVec$ is in the column space of $\tildeXMat$, and
\begin{align*}
  \E[\XMat\betaAgLS]-\XMat\betaTrueVec&=\XMat(\AMat\AMat^+-{\bf
    I}_p)\betaTrueVec+\tildeXMat\tildeXMat^{+}\XMat({\bf
    I}_p-\AMat\AMat^+)\betaTrueVec\\
&=(\tildeXMat\tildeXMat^{+}-{\bf
    I}_n)\XMat({\bf I}_p-\AMat\AMat^+)\betaTrueVec
\end{align*}
\begin{align*}
\E\|\XMat\betaAgLS-\XMat\betaTrueVec\|^2&=\|(\tildeXMat\tildeXMat^{+}-{\bf
    I}_n)\XMat({\bf I}_p-\AMat\AMat^+)\betaTrueVec \|^2+ \sigma^2
  \text{tr}(\tildeXMat[\tildeXMat^T \tildeXMat]^{-1}\tildeXMat^T)\\
&=\|(\tildeXMat\tildeXMat^{+}-{\bf
    I}_n)\XMat({\bf I}_p-\AMat\AMat^+)\betaTrueVec \|^2+
  \sigma^2|\mathcal C|\\
&\le \|\XMat\|_{op}^2 \|({\bf I}_p-\AMat\AMat^+)\betaTrueVec \|^2+
  \sigma^2|\mathcal C|
\end{align*}
since $\|(\tildeXMat\tildeXMat^{+}-{\bf I}_n)\|_{op}=1$ and the trace of a projection matrix is the rank of the target space.  Finally,
observe that 
$$
\|({\bf I}_p-\AMat\AMat^+)\betaTrueVec \|^2=\sum_{\ell=1}^{|\mathcal
    C|}\sum_{j\in\mathcal C_\ell}\left(\beta^*_j - |\mathcal
    C_\ell|^{-1}\sum_{j'\in \mathcal C_\ell}\beta^*_{j'}
  \right)^2.
$$


\section{Consensus ADMM for Solving Problem \eqref{ourapproach}}
\label{sec:admm}

The ADMM algorithm is given in Algorithm \ref{admm}. 
Let $\XMat=\text{SVD}_{\text{compact}}({\widetilde \UMat, \widetilde \DMat, \widetilde \VMat})$ be the {\em compact} singular value decomposition of $\XMat$, where $\widetilde\DMat \in \real^{\min(n, p) \times \min(n, p)}$ is a diagonal matrix with non-zero singular values on the diagonal, and $\widetilde \UMat\in\real^{n\times \min(n, p)}$ and $\widetilde \VMat\in \real^{p\times \min(n, p)}$ contain the left and right singular vectors in columns corresponding to non-zero singular values, respectively. Similarly, we have $(\IpMat:-\bm{A})=\text{SVD}_{\text{compact}}(\boldsymbol \cdot, \boldsymbol \cdot, \widetilde \QMat)$ where $\widetilde \QMat\in \real^{(p+|\T|):p}$ contains $p$ right singular vectors correponding to non-zero singular values. 

\begin{algorithm}
\spacingset{1.2}
\caption{\em Consensus ADMM for Solving Problem \eqref{ourapproach}}
\begin{algorithmic}[1]
\Input  $\yVec, \XMat, \AMat, n, p, |\T|, \lambda, \alpha, \rho, \epsilon^{abs}, \epsilon^{rel}, \texttt{maxite}$.
\State $\XMat = \text{SVD}_{\text{compact}}(\boldsymbol \cdot, \widetilde \DMat, \widetilde \VMat)$
\State $(\IpMat:-\AMat)=\text{SVD}_{\text{compact}}(\boldsymbol \cdot, \boldsymbol \cdot, \widetilde \QMat)$
\State $\bm{\beta}^{0} \leftarrow \bm{\beta}^{(i)0} \leftarrow \bm{v}^{(i)0}\leftarrow {\bm 0}\in\mathbb R^{p}\quad\forall i = 1,2,3$
\State $\bm{\gamma}^{0}\leftarrow \bm{\gamma}^{(j)0} \leftarrow \bm{u}^{(j)0}\leftarrow {\bm 0}\in\mathbb R^{|\T|}\quad\forall j = 1,2$
\State $\texttt{continue}\leftarrow \texttt{true}$
\State $k\leftarrow 0$
\While{$k < \texttt{maxite}$\AND \texttt{continue}}
\State $k\leftarrow k +1$
\State $\betaOneVecK \leftarrow 
\left[\widetilde \VMat diag\left( \frac{1}{[\widetilde \DMat^T\widetilde \DMat]_{ii} + n\rho} \right)\widetilde \VMat^T + \frac{1}{n\rho}(\IpMat - \widetilde \VMat \widetilde \VMat^T) \right]\left(\XMat^T\yVec + n\rho\betaVec^{k-1} - n\vOneVecKMinus \right)$
\State $\beta^{(2)k}_j \leftarrow S\left(\beta^{k-1}_j -
  \frac{1}{\rho}v^{(2)k-1}_j, \frac{\lambda(1-\alpha)\tilde w_j}{\rho}
\right)\quad\forall j=1, \ldots, p$
\State $\gamma^{(1)k}_\ell \leftarrow
\begin{cases}S\left(\gamma^{k-1}_\ell -
    \frac{1}{\rho}u^{(1)k-1}_\ell, \frac{\lambda\alpha w_\ell}{\rho}
  \right) & \text{if } w_\ell>0\\
\gamma^{k-1}_\ell - \frac{1}{\rho}u^{(1)k-1}_\ell & \text{if }w_\ell =0  \end{cases}$
\State $\begin{pmatrix}\betaThreeVecK\\\gammaTwoVecK  \end{pmatrix}\leftarrow \left( \bm{I}_{p+|\T|} - \widetilde \QMat\widetilde \QMat^T \right)\left[ \begin{pmatrix}\betaVec^{k-1}\\\gammaVec^{k-1} \end{pmatrix} - \frac{1}{\rho}\begin{pmatrix}\vThreeVecKMinus\\ \uTwoVecKMinus  \end{pmatrix}     \right]$
\State $\betaVec^{k} \leftarrow {(\betaOneVecK +\betaTwoVecK + \betaThreeVecK)}/3$
\State $\gammaVec^{k} \leftarrow (\gammaOneVecK + \gammaTwoVecK)/2$ 
\State $\vIVecK \leftarrow \vIVecKMinus + \rho (\betaIVecK - \betaVec^k)\quad\forall i = 1, 2, 3$
\State $\uJVecK \leftarrow \uJVecKMinus + \rho(\gammaJVecK - \gammaVec^k)\quad\forall j = 1, 2.
$
\IfNoThen{{\tiny$\sqrt{\sum_{i=1}^3\|\betaVec^{(i)k} - \betaVec^k\|_2^2 + \sum_{j=1}^2 \left\|\gammaVec^{(j)k} - \gammaVec^k\right\|_2^2}
\le  \epsilon^{abs}\sqrt{3p + 2|\T|} + \epsilon^{rel}  \max\left\{\sqrt{\sum_{i=1}^3\|\betaVec^{(i)k}\|_2^2 + \sum_{j=1}^2 \left\|\gammaVec^{(j)k}\right\|_2^2},  \sqrt{3\left\| \betaVec^k\right\|_2^2 + 2 \left\| \gammaVec^k\right\|_2^2}  \right\}$}}
\StatexIndent[0.7] \algorithmicand {\tiny $\rho \sqrt{3 \| \betaVec^k - \betaVec^{k-1} \|_2^2 + 2\|\gammaVec^k - \gammaVec^{k-1}\|_2^2}
\le  \epsilon^{abs} \sqrt{3p + 2|\T|} + \epsilon^{rel}  \sqrt{\sum_{i=1}^3\|\vVec^{(i)k}\|_2^2 + \sum_{j=1}^2 \|\uVec^{(j)k}\|_2^2}$} 
\algorithmicthen
\State $\texttt{continue}\leftarrow \texttt{false}$
\EndIf
\EndWhile
\Output $\betaVec^{k}, \gammaVec^{k}$
\end{algorithmic}
\label{admm}
\end{algorithm}

\subsection{Derivation of Algorithm \ref{admm}}
\label{appenADMM}
The ADMM updates involve minimizing the augmented Lagrangian of the global consensus problem \eqref{consensus},
\begin{align*}
&L_{\rho}(\betaOneVec, \betaTwoVec, \betaThreeVec, \gammaOneVec, \gammaTwoVec, \betaVec, \gammaVec; \vOneVec, \vTwoVec, \vThreeVec, \uOneVec, \uTwoVec)\\
=\ & \frac{1}{2n}\left\|\yVec-\XMat\betaOneVec\right\|_2^2 +\lambda\left( \alpha\sum_{\ell = 1}^{|\T|}w_\ell \left| \gamma_\ell^{(1)}\right| +(1-\alpha)\sum_{j=1}^p\tilde w_j \left| \beta_j^{(2)} \right|  \right) + 1_\infty\{\betaThreeVec = A\gammaTwoVec\}\nonumber\\
&+ \sum_{i=1}^3 \left(\bm{v}^{(i)T}(\bm\beta^{(i)} - \betaVec) + \frac{\rho}{2}\|\bm\beta^{(i)}-\betaVec\|_2^2  \right)
+ \sum_{j=1}^2\left(\bm u^{(j)T}(\bm{\gamma}^{(j)} - \gammaVec) +\frac{\rho}{2}\|\bm{\gamma}^{(j)}-\gammaVec\|_2^2 \right).
\end{align*}

\begin{enumerate}
\item Update $\betaOneVec$.
$$
\betaOneVecKPlus :=\argmin_{\betaOneVec\in\real^p}\left\{\frac{1}{2n}\left\| \yVec - \XMat\betaOneVec \right\|_2^2+ \left<\vOneVecK, (\betaOneVec - \betaVec^k) \right>+ \frac{\rho}{2}\|\betaOneVec-\betaVec^k\|_2^2  \right\}.
$$
Let $\XMat=\text{SVD}(\UMat,  \DMat,  \VMat)$ be the singular value decomposition of $\XMat$, where $\UMat\in\real^{n\times n}$ contains left singular vectors in columns, $\VMat\in \real^{p\times p}$ contains right singular vectors in columns, and $\DMat\in\real ^{n\times p}$ is a rectangular diagonal matrix with decreasing singular values on the diagonal. First order condition to the above problem gives us:
\begin{align*}
&(\XMat^T\XMat + n\rho \IpMat)\betaOneVecKPlus = \XMat^T\yVec + n\rho\betaVec^k - n\vOneVecK\\
\Rightarrow\ & \VMat(\DMat^T\DMat+n\rho \IpMat)\VMat^T\betaOneVecKPlus = \XMat^T\yVec + n\rho\betaVec^k - n\vOneVecK\\
\Rightarrow\ & \betaOneVecKPlus = \VMat diag\left( ({[\DMat^T\DMat]_{ii} + n\rho})^{-1} \right)\VMat^T\left(\XMat^T\yVec + n\rho\betaVec^k - n\vOneVecK \right).
\end{align*}
When $n\ge p$,  we have 
\begin{equation}\label{beta1_case1}
\betaOneVecKPlus = \widetilde \VMat diag\left( ({[\widetilde \DMat^T\widetilde \DMat]_{ii} + n\rho})^{-1} \right)\widetilde \VMat^T\left(\XMat^T\yVec + n\rho\betaVec^k - n\vOneVecK \right).
\end{equation}
When $n<p$, the SVD can be expressed in a compact form: $\DMat=(\widetilde \DMat:\bm 0)$ and $\VMat= (\widetilde \VMat:\widetilde \VMat_\perp)$ where $\widetilde \DMat\in \real^{n\times n}$ and $\widetilde \VMat\in \real^{p\times n}$ are from the compact SVD of $\XMat$, and $\widetilde \VMat_\perp\in\real^{p\times (p-n)}$. Thus,
\begin{align*}
\VMat diag\left( ({[\DMat^T\DMat]_{ii} + n\rho})^{-1} \right)\VMat^T
&=\begin{pmatrix} \widetilde \VMat:\widetilde \VMat_\perp \end{pmatrix} diag\left( ({[\DMat^T\DMat]_{ii} + n\rho})^{-1} \right) \begin{pmatrix} \widetilde \VMat^T\\ \widetilde \VMat^T_\perp  \end{pmatrix}\\
&= \widetilde \VMat diag\left( ({[\widetilde \DMat^T\widetilde \DMat]_{ii} + n\rho})^{-1} \right) \widetilde \VMat^T + \widetilde \VMat_\perp \widetilde \VMat^T_\perp/(n\rho)\\
&= \widetilde \VMat diag\left( ({[\widetilde \DMat^T\widetilde \DMat]_{ii} + n\rho})^{-1} \right) \widetilde \VMat^T + (\IpMat - \widetilde \VMat\widetilde \VMat^T)/(n\rho).
\end{align*}
So when $n < p$,
\begin{equation}\label{beta1_case2}
\betaOneVecKPlus = \left[\widetilde \VMat diag\left( ({[\widetilde \DMat^T\widetilde \DMat]_{ii} + n\rho})^{-1} \right)\widetilde \VMat^T  + (\IpMat - \widetilde \VMat\widetilde \VMat^T)/(n\rho)\right]\left(\XMat^T\yVec + n\rho\betaVec^{k} - n\vOneVecK \right).
\end{equation}
Since $\widetilde \VMat = \VMat$ when $n \ge p$ and $\VMat \VMat^T = \IpMat$, we have \eqref{beta1_case2} boil  to \eqref{beta1_case1} in that case.

\item Update $\betaTwoVec$.
$$
\betaTwoVecKPlus := \argmin_{\betaTwoVec\in\real^p}
\left\{\frac{\rho}{2} \left\|\betaTwoVec - \left(\betaVec^k -
      \frac{1}{\rho}\vTwoVecK\right)  \right\|_2^2  + \lambda
  (1-\alpha)\sum_{j=1}^p\tilde w_j \left| \beta_j^{(2)}
  \right|\right\}. 
$$
The solution is simply elementwise soft-thresholding:
$$
\beta^{(2)k+1}_j =S\left(\beta^k_j - \frac{1}{\rho}v^{(2)}_j, \frac{\lambda(1-\alpha)\tilde w_j }{\rho}  \right)\quad\forall j=1, \ldots, p.
$$

\item Update $\gammaOneVec$.
$$
\gammaOneVecKPlus := \argmin_{\gammaOneVec\in\real^{|\T|}} \left\{\frac{\rho}{2} \left\|\gammaOneVec - \left(\gammaVec^k - \frac{1}{\rho}\uOneVecK\right)  \right\|_2^2  + \lambda \alpha\sum_{\ell = 1}^{|\T|}w_\ell \left| \gamma_\ell^{(1)}\right|\right\}.
$$
Since sometimes we will choose to $w_\ell=0$ for the root, we break
the solution into cases:
$$
\gamma^{(1)k+1}_\ell = \begin{cases}S\left(\gamma^k_\ell -
    \frac{1}{\rho}u^{(1)k}_\ell, \frac{\lambda\alpha w_\ell}{\rho}
  \right) & \text{if } w_\ell>0\\
\gamma^k_\ell - \frac{1}{\rho}u^{(1)k}_\ell & \text{if }w_\ell =0.  \end{cases}
$$
\item Joint update of $\betaThreeVec$ and $\gammaTwoVec$.
\begin{align*}
\begin{pmatrix}\betaThreeVecKPlus\\\gammaTwoVecKPlus  \end{pmatrix}
:= &\argmin_{\betaThreeVec\in\real^p, \gammaTwoVec\in\real^{|\T|}}
\left\{\left\|\betaThreeVec - \left(\betaVec^k - \frac{1}{\rho}\vThreeVecK\right)  \right\|_2^2 + \left\| \gammaTwoVec - \left(\gammaVec^k - \frac{1}{\rho}\uTwoVecK\right) \right\|_2^2  \right\}\\
&\text{s.t. }(\IpMat: -\AMat)\begin{pmatrix}\betaThreeVec\\\gammaTwoVec  \end{pmatrix}= 0.
\end{align*}
The solution is the projection of $\begin{pmatrix}\betaVec^k\\\gammaVec^k \end{pmatrix} - \frac{1}{\rho}\begin{pmatrix}\vThreeVecK\\ \uTwoVecK  \end{pmatrix} $ onto the null space of $(\IpMat: -\AMat)$. Let $(\IpMat:-\AMat)=\text{SVD}(\boldsymbol \cdot,\boldsymbol, \QMat)$ where $\QMat =(\widetilde \QMat: \widetilde \QMat_\perp)\in \real^{(p+|\T|):(p+|\T|)}$ contains all the right singular vectors in columns. So $\bm{I}_{p+|\T|}=\QMat \QMat^T = \widetilde \QMat \widetilde \QMat^T + \widetilde \QMat_\perp \widetilde \QMat_\perp^T$. Since $\widetilde \QMat$ corresponds to non-zero singular values of $(\IpMat:-\AMat)$ by construction, we have $\widetilde \QMat_\perp$ corresponds to the zero singular values, making itself an orthonormal basis for the null space of $(\IpMat:-\AMat)$. Thus, 
\begin{align*}
\begin{pmatrix}\betaThreeVecKPlus\\\gammaTwoVecKPlus  \end{pmatrix} &= \widetilde \QMat_\perp (\widetilde \QMat_\perp^T \widetilde \QMat_\perp)^{-1} \widetilde \QMat_\perp^T\left[ \begin{pmatrix}\betaVec^k\\\gammaVec^k \end{pmatrix} - \frac{1}{\rho}\begin{pmatrix}\vThreeVecK\\ \uTwoVecK  \end{pmatrix}     \right]\\
&= \widetilde \QMat_\perp \widetilde \QMat_\perp^T \left[ \begin{pmatrix}\betaVec^k\\\gammaVec^k \end{pmatrix} - \frac{1}{\rho}\begin{pmatrix}\vThreeVecK\\ \uTwoVecK  \end{pmatrix}     \right]\\
&= \left(\bm{I}_{p+|\T|} - \widetilde \QMat \widetilde \QMat^T\right) \left[ \begin{pmatrix}\betaVec^k\\\gammaVec^k \end{pmatrix} - \frac{1}{\rho}\begin{pmatrix}\vThreeVecK\\ \uTwoVecK  \end{pmatrix}     \right]
\end{align*}

\item Update global variables $\betaVec$ and $\gammaVec$.
\begin{align}
\betaVec^{k+1} & := \argmin_{\betaVec\in \real^p}\sum_{i=1}^3 \left\| \betaVec - \left(\betaIVecKPlus + \frac{1}{\rho}\vIVecK\right) \right\|_2^2  = \bar\betaVec^{k+1} + \frac{1}{\rho}\bar \vVec^{k}\label{global1}\\
\gammaVec^{k+1} & := \argmin_{\gammaVec\in \real^{|\T|}}\sum_{j=1}^2\left\|\gammaVec - \left(\gammaJVecKPlus + \frac{1}{\rho}\uJVecK\right)  \right\|_2^2 = \bar\gammaVec^{k+1} + \frac{1}{\rho}\bar \uVec^{k}\label{global2}
\end{align}
where $\bar\betaVec^k := \frac{\betaOneVecK + \betaTwoVecK +\betaThreeVecK}{3}$, $\bar \vVec^k := \frac{\vOneVecK + \vTwoVecK + \vThreeVecK}{3}$, $\bar\gammaVec^k := \frac{\gammaOneVecK + \gammaTwoVecK}{2} $ and $\bar \uVec^k := \frac{\uOneVecK + \uTwoVecK}{2} $.

\item Update dual variables.
\begin{align*}
\vOneVecKPlus & := \vIVecK + \rho (\betaIVecKPlus - \betaVec^{k+1})\quad\text{for }i = 1, 2,3,\\
\uOneVecKPlus & := \uJVecK + \rho (\gammaJVecKPlus - \gammaVec^{k+1})\quad\text{for }j=1,2.
\end{align*}
Similarly, averaging the updates for $u$ and the udpates for $v$ gives
\begin{align}
\bar \vVec^{k+1} & = \bar \vVec^k + \rho (\bar \betaVec^{k+1} - \betaVec^{k+1})\label{multiplier1}\\
\bar \uVec^{k+1} & = \bar \uVec^k + \rho (\bar\gammaVec^{k+1}-\gammaVec^{k+1})\label{multiplier2}
\end{align}
Substituting \eqref{global1} and \eqref{global2} into \eqref{multiplier1} and \eqref{multiplier2} yields that $\bar \vVec^{k+1}=\bar \uVec^{k+1}=0$ after the first iteration. 
\end{enumerate}

Using $\betaVec^k = \bar \betaVec^k$ and $\gammaVec^k = \bar\gammaVec^k$ in the above updates, the updates become Lines 9-16 of Algorithm \ref{admm}. Next, we follow Section 3.3.1 in \cite{boyd2011} to determine the termination criteria. We first write Problem \eqref{consensus} in the same form as Problem (3.1) in \cite{boyd2011} which is presented below in typewriter font:
\begin{equation*}
\mathtt{\min_{x, z} \quad\{f(x) + g(z) \text{ s.t. } Ax + Bz = c \}}
\end{equation*}
where 
\begin{equation*}
\mathtt{A} = \bm{I}_{3p + 2|\T|}, \mathtt{B} = -\begin{pmatrix}\IpMat & 0\\ \IpMat & 0\\ \IpMat & 0\\0 & \ITMat  \\ 0 & \ITMat \end{pmatrix}, \mathtt{c} = 0, \mathtt{x} = \begin{pmatrix} \betaOneVec \\\betaTwoVec\\\betaThreeVec\\\gammaOneVec\\\gammaTwoVec \end{pmatrix}\text{ and } \mathtt{z} = \begin{pmatrix} \betaVec\\\gammaVec \end{pmatrix}.
\end{equation*}
The primal and dual residuals are 
\begin{equation*}
\mathtt r^{k+1} = \mathtt A \mathtt x^{k+1} + \mathtt B \mathtt z^{k+1} - c
= \begin{pmatrix}\betaOneVecKPlus - \betaVec^{k+1}\\\betaTwoVecKPlus - \betaVec^{k+1} \\\betaThreeVecKPlus - \betaVec^{k+1}\\\gammaOneVecKPlus - \gammaVec^{k+1}\\\gammaTwoVecKPlus - \gammaVec^{k+1} \end{pmatrix}\text{ and }
\mathtt s^{k+1} = \rho \mathtt A^T \mathtt B (\mathtt z^{k+1} - \mathtt z^k) =  
\rho\begin{pmatrix} \betaVec^{k+1} - \betaVec^k\\ \betaVec^{k+1} - \betaVec^k \\ \betaVec^{k+1} - \betaVec^k \\ \gammaVec^{k+1} - \gammaVec^k \\ \gammaVec^{k+1} - \gammaVec^k  \end{pmatrix}.
\end{equation*}

By Condition (3.12) in \cite{boyd2011}, the ADMM algorithm stops when both residuals are small. In our case, the termination criteria are the following.
\begin{enumerate}
\item The primal residual is small: 
\begin{align*}
&\sqrt{\sum_{i=1}^3\|\betaIVecK - \betaVec^k\|_2^2 + \sum_{j=1}^2 \left\|\gammaJVecK - \gammaVec^k\right\|_2^2}\\
\le & \sqrt{3p + 2|\T|}\cdot\epsilon^{abs} + \epsilon^{rel} \cdot \max\left\{\sqrt{\sum_{i=1}^3\|\betaIVecK\|_2^2 + \sum_{j=1}^2 \left\|\gammaJVecK\right\|_2^2},  \sqrt{3\left\| \betaVec^k\right\|_2^2 + 2 \left\| \gammaVec^k\right\|_2^2}  \right\}.
\end{align*}
\item The dual residual is small:
\begin{equation*}
\rho \cdot \sqrt{3 \| \betaVec^k - \betaVec^{k-1} \|_2^2 + 2\|\gammaVec^k - \gammaVec^{k-1}\|_2^2}
\le  \sqrt{3p + 2|\T|} \cdot \epsilon^{abs} + \epsilon^{rel} \cdot \sqrt{\sum_{i=1}^3\|\vIVecK\|_2^2 + \sum_{j=1}^2 \|\uJVecK\|_2^2}.
\end{equation*}
\end{enumerate}

\subsection{Treatment of Intercept in Problem \eqref{ourapproach}}
\label{ourapproach_intercept}

When an intercept $\beta_0$ is included in the least squares, Problem \eqref{ourapproach} becomes:
\begin{equation}\label{ourapproach_var2}
\min_{\substack{\beta_0\in \real, \betaVec\in \real^p, \gammaVec\in\real^{|\T|}\\\st{\betaVec=\AMat\gammaVec}}}\left\{ \frac{1}{2n} \left\| \yVec - \XMat\betaVec - \beta_0 \bm {1}_n \right\|_2^2 +\lambda\left( \alpha\sum_{\ell = 1}^{|\T|}w_\ell \left| \gamma_\ell\right| +(1-\alpha)\sum_{j=1}^p\tilde w_j \left| \beta_j \right|  \right)   \right\}.
\end{equation}
First-order coniditon of the solution $({\hat\beta}_0, \hat\betaVec)$ yields that
\begin{equation*}
\frac{\partial\frac{1}{2n} \left\| \yVec - \XMat\betaVec - \beta_0 \bm {1}_n\right\|_2^2}{\partial \beta_0}\Bigg|_{(\beta_0, \betaVec)=({\hat\beta}_0, \hat\betaVec)}
= \frac{1}{n}\bm {1}_n^T(\bm {1}_n{\hat\beta}_0 - (\yVec - \XMat \hat\betaVec)) = \frac{1}{n} (n{\hat\beta}_0 - \bm {1}_n^T(\yVec-\XMat\hat\betaVec)) = 0.
\end{equation*}
So ${\hat\beta}_0 = \frac{1}{n} \bm {1}_n^T(\yVec -
\XMat\hat\betaVec)$. Plugging ${\hat\beta}_0$ in Problem
\eqref{ourapproach_var2} and letting ${\bf H}=\InMat - \frac{1}{n}\bm {1}_n\bm {1}_n^T$ yields
\begin{equation*}
\min_{\substack{\betaVec\in \real^p,
    \gammaVec\in\real^{|\T|}\\\st{\betaVec=\AMat\gammaVec}}}\left\{
  \frac{1}{2n} \left\| {\bf H}\yVec - {\bf H}\XMat\betaVec \right\|_2^2 +\lambda\left( \alpha\sum_{\ell = 1}^{|\T|}w_\ell \left| \gamma_\ell\right| +(1-\alpha)\sum_{j=1}^p\tilde w_j \left| \beta_j \right|  \right)    \right\},
\end{equation*}
which can now be solved using our consensus ADMM algorithm.


\section{Proof of Lemma \ref{lem_aggregating_set}}
\label{proof_lem_aggregating_set}

We first show existence of such $B^*$ by providing a feasible procedure to find $B^*$. Suppose $\betaTrueVec$ has at least two distinct values (otherwise $B^*=\{r\}$ trivially). Start with $B=\mathcal L(\T)$ so that the first constraint is satisfied. If for siblings $u, v$ in $B$ such that the second constraint is violated, by construction $\beta^*_j=\beta^*_k$ for $j\in \mathcal L(\T_u)$ and $k\in \mathcal L(\T_v)$. So we replace $u, v$ in $B$ with their parent node. We repeat the above steps until the second constraint is satisfied, while holding the first constraint. Thus, $B$ satisfies the two requirements for $B^*$.

Suppose $B^*$ and $\tilde B^*$ are different aggregating sets for $\betaTrueVec$. Without loss of generality, suppose there exists $u\in \tilde B^*$ but $u\notin B^*$. Then $u$ is a descendant or an ancestor of some nodes in $B^*$; for either case the second constraint will be violated. Thus, such $u$ does not exist and $\tilde B^*=B^*$.

The existence and uniqueness of $\mathcal A^*$ follow from the definition of support of $\betaTrueVec$.


\section{Proof of Theorem \ref{maintheorem_alpha2}}
\label{proof_maintheorem_alpha2}

Denote our penalty by $\Omega(\betaVec, \gammaVec) := \alpha\sum_{\ell = 1}^{|\T|}w_\ell \left| \gamma_\ell\right| +(1-\alpha)\sum_{j=1}^p\tilde w_j \left| \beta_j \right|$. We follow the proof strategy used in Theorem 1 of \cite{lou2016} to prove this theorem. If $(\hat\betaVec, \hat\gammaVec)$ is a solution to Problem \eqref{ourapproach}, then we have
$$
\frac{1}{2n}\left\| \yVec-\XMat\betaOurVec \right\|_2^2 + \lambda\Omega(\betaOurVec, \gammaOurVec) \le 
\frac{1}{2n}\left\| \yVec-\XMat\betaVec \right\|_2^2 + \lambda\Omega(\betaVec, \gammaVec)
$$
for any $(\betaVec, \gammaVec)$ such that $\betaVec = \AMat\gammaVec$. Let $(\betaTrueVec, \gammaTrueVec)$ be such that 
$$
\betaTrueVec = \AMat_{B^*}\tildebetaTrueVec\quad\text{and}\quad \gamma^*_\ell = \begin{cases}\tilde \beta^*_\ell&\text{ if }\ell\in B^*\\0&\text{ otherwise.} \end{cases}
$$
Plugging in $\yVec=\XMat\betaTrueVec +\epsilonVec$ and $(\betaVec, \gammaVec) = (\betaTrueVec, \gammaTrueVec)$, with some algebra we have
\begin{equation}\label{temp_appenf_1}
\frac{1}{2n}\left\| \XMat\betaOurVec -\XMat\betaTrueVec\right\|_2^2 + \lambda\Omega(\betaOurVec, \gammaOurVec) \le 
\lambda\Omega(\betaTrueVec, \gammaTrueVec)+\frac{1}{n}\epsilonVec^T\XMat\hat\DeltaVec^{(\betaTrueVec)}
\end{equation}
where $\hat\DeltaVec^{(\betaTrueVec)}=\betaOurVec - \betaTrueVec$. By $\betaOurVec = \AMat\gammaOurVec$ and $\betaTrueVec= \AMat\gammaTrueVec$ (and writing $\hat\DeltaVec^{(\gammaTrueVec)}=\gammaOurVec - \gammaTrueVec$),
$$
\frac{1}{n}\epsilonVec^T\XMat\hat\DeltaVec^{(\betaTrueVec)} = \frac{1}{n}\epsilonVec^T\XMat \AMat\hat\DeltaVec^{(\gammaTrueVec)}.
$$
We next bound $n^{-1}\epsilonVec^T\XMat\hat\DeltaVec^{(\betaTrueVec)}= (1-\alpha)n^{-1}\epsilonVec^T\XMat\hat\DeltaVec^{(\betaTrueVec)}+ \alpha n^{-1}\epsilonVec^T\XMat \AMat\hat\DeltaVec^{(\gammaTrueVec)}$ in absolute value. Define $V_j:=n^{-1/2}\XMat_j^T\epsilonVec$ for $j=1, \ldots, p$ and $U_\ell:=n^{-1/2}\AMat_\ell^T\XMat^T\epsilonVec$ for $\ell=1, \ldots, |\T|$. With the choice of weights $\tilde w_j = \left\| \XMat_j \right\|_2/\sqrt{n}$ for $1\le j\le p$ and $w_\ell = \left\| \XMat \AMat_\ell \right\|_2/\sqrt{n}$ for $1\le \ell \le |\T|$, 
\begin{align}\label{temp_appenf_2}
\left|\frac{1}{n}\epsilonVec^T\XMat\hat\DeltaVec^{(\betaTrueVec)}\right|
&=\left| (1- \alpha) \frac{1}{n}\epsilonVec^T \XMat\hat\DeltaVec^{(\betaTrueVec)} + \alpha \frac{1}{n}\epsilonVec^T \XMat \AMat\hat\DeltaVec^{(\gammaTrueVec)} \right|\nonumber\\
&= \left|(1-\alpha) \sum_{j=1}^p \left\{\frac{\XMat_j^T\epsilonVec}{\sqrt{n}\|\XMat_j\|_2} \cdot \frac{\|\XMat_j\|_2}{\sqrt{n}} \hat\Delta_j^{(\betaTrueVec)} \right\} 
+ \alpha \sum_{\ell=1}^{|\T|} \left\{\frac{\AMat_\ell^T\XMat^T\epsilonVec}{\sqrt{n}\|\XMat\AMat_\ell\|_2} \cdot \frac{\|\XMat\AMat_\ell\|_2}{\sqrt{n}} \hat\Delta_\ell^{(\gammaTrueVec)} \right\} \right| \nonumber\\
&=\left| (1-\alpha) \sum_{j=1}^p\left\{\frac{V_j}{\|\XMat_j\|_2}\cdot\tilde w_j \hat \Delta_j^{(\betaTrueVec)} \right\} 
+ \alpha \sum_{\ell=1}^{|\T|} \left\{ \frac{U_\ell}{\|\XMat\AMat_\ell\|_2}\cdot w_\ell \hat\Delta^{(\gammaTrueVec)}_\ell \right\}  \right|\nonumber \\
&\le(1-\alpha) \sum_{j=1}^p\left\{\frac{|V_j|}{\|\XMat_j\|_2}\cdot\tilde w_j \left| \hat \Delta_j^{(\betaTrueVec)} \right| \right\}
+ \alpha \sum_{\ell=1}^{|\T|} \left\{ \frac{|U_\ell|}{\|\XMat\AMat_\ell\|_2}\cdot w_\ell \left| \hat\Delta^{(\gammaTrueVec)}_\ell \right| \right\}\nonumber\\
&\le(1-\alpha) \max_{1\le j\le p} \frac{|V_j|}{\|\XMat_j\|_2}\cdot \sum_{j=1}^p \tilde w_j \left| \hat \Delta_j^{(\betaTrueVec)} \right|
+ \alpha \max_{1\le \ell \le |\T|} \frac{|U_\ell|}{\|\XMat\AMat_\ell\|_2}\cdot  \sum_{\ell=1}^{|\T|} w_\ell \left| \hat\Delta^{(\gammaTrueVec)}_\ell \right|
\end{align}
where the second to last inequality follows from the triangle inequality. 

Since $\epsilonVec\sim N_n(0, \sigma^2 \InMat)$, $V_j\sim N\left(0, \|\XMat_j\|_2^2\sigma^2/n \right)$ for $j=1, \ldots, p$ and $U_\ell\sim N\left(0, \|\XMat \AMat_\ell\|_2^2\sigma^2/n  \right)$ for $\ell=1, \ldots, |\T|$. By Lemma 6.2 of \cite{bhlmann2011}, we have for $x>0$
\begin{align*}
&\mathbb P\left(\max_{1\le j\le p} \frac{|V_j|}{\|\XMat_j\|_2}>  \sigma\sqrt{\frac{2(x+\log p)}{n}}   \right)\le 2e^{-x}\quad\text{and}\\
&\mathbb P\left(\max_{1\le\ell\le |\T|} \frac{|U_\ell|}{\|\XMat\AMat_\ell\|_2}>  \sigma\sqrt{\frac{2(x+\log |\T|)}{n}}   \right)\le 2e^{-x}.
\end{align*}
By the union bound,
$$
\mathbb P\left(\max_{1\le j\le p} \frac{|V_j|}{\|\XMat_j\|_2}>  \sigma\sqrt{\frac{2(x+\log p)}{n}}\quad\text{or}\quad
\max_{1\le\ell\le |\T|} \frac{|U_\ell|}{\|\XMat\AMat_\ell\|_2}>  \sigma\sqrt{\frac{2(x+\log |\T|)}{n}}   \right)\le 4e^{-x}.
$$ 
By the construction of $\T$, each internal node has at least 2 child
nodes. To go up to the next level from the leaf nodes, only one node
``survives'' among its siblings. For $\T$ with $p$ leaf nodes, there
must be at most $p-1$ internal nodes where the maximum number is
achieved when $\T$ is a full binary tree. So $|\T|\le 2p$. Choosing $x
= \log (2p)$, we have that the following results hold with probability at least $1-2/p$:
$$
\max_{1\le j\le p} \frac{|V_j|}{\|\XMat_j\|_2} \le 2\sigma\sqrt{\frac{\log (2p)}{n}}\quad\text{and}\quad
\max_{1\le\ell\le |\T|} \frac{|U_\ell|}{\|\XMat\AMat_\ell\|_2} \le 2\sigma\sqrt{\frac{\log (2p)}{n}}.
$$
Plugging these upper bounds into \eqref{temp_appenf_2}, we have the following inequality holding with high probability:
\begin{align}\label{temp_appenf_4}
\left|\frac{1}{n}\epsilonVec^T\XMat\hat\DeltaVec^{(\betaTrueVec)}\right|
&\le 
2\sigma\sqrt{\frac{\log(2p)}{n}}\left( (1-\alpha)\sum_{j=1}^p \tilde w_j \left|\hat\Delta_j^{(\betaTrueVec)} \right| + \alpha\sum_{\ell=1}^{|\T|}w_\ell \left|\hat\Delta_\ell^{(\gammaTrueVec)} \right|\right)\nonumber\\
&=2\sigma\sqrt{\frac{\log(2p)}{n}}\Omega(\hat\DeltaVec^{(\betaTrueVec)}, \hat\DeltaVec^{(\gammaTrueVec)})
\end{align}

Let $\lambda \ge 4\sigma\sqrt{\log(2p)/n}$ and $0\le \alpha\le 1$. By \eqref{temp_appenf_1} and \eqref{temp_appenf_4}, the following holds with probability at least $1-2/p$:
\begin{align*}
\frac{1}{2n}\left\| \XMat\betaOurVec -\XMat\betaTrueVec\right\|_2^2
&\le \frac{1}{2}\lambda\Omega(\hat\DeltaVec^{(\betaTrueVec)}, \hat\DeltaVec^{(\gammaTrueVec)}) - \lambda\Omega(\betaOurVec, \gammaOurVec) + \lambda\Omega(\betaTrueVec, \gammaTrueVec)\\
&\le \frac{1}{2}\left( \lambda\Omega(\betaOurVec, \gammaOurVec) + \lambda\Omega(\betaTrueVec, \gammaTrueVec) \right) - \lambda\Omega(\betaOurVec, \gammaOurVec) + \lambda\Omega(\betaTrueVec, \gammaTrueVec)\quad (\text{by triangle inequality})\\
&\le \frac{3}{2}\lambda\Omega(\betaTrueVec, \gammaTrueVec)
=\frac{3}{2}\lambda \left( (1-\alpha)\sum_{j=1}^p \tilde w_j |\beta^*_j| + \alpha\sum_{\ell=1}^{|\T|}w_\ell |\gamma^*_\ell|  \right)\\
&= \frac{3}{2}\lambda \left( (1-\alpha)\sum_{j\in\mathcal A^*} \tilde w_j |\beta^*_j| + \alpha\sum_{\ell\in B^*}w_\ell |\tilde\beta^*_\ell|  \right)
\end{align*}
where $\mathcal A^*$ is the support of $\betaTrueVec$.


\section{Proof of Corollary \ref{cor_lasso_rate}}
\label{sec:proof_of_cor_lasso_rate}

With $\|\XMat_j\|_2\le\sqrt{n}$ the weights in Theorem \ref{maintheorem_alpha2} become:
\begin{itemize}
\item For $1\le j \le  p$, $\tilde w_j  = \frac{\|\XMat_j\|_2}{\sqrt{n}} \le 1$.
\item For $1\le \ell \le  |\T|$, $w_\ell = \frac{\|\XMat \AMat_\ell\|_2}{\sqrt{n}} \le \frac{1}{\sqrt{n}}\sum_{j \in \mathcal L(\mathcal T_{u_\ell})}\|\XMat_j\|_2\le \left|\mathcal L(\mathcal T_{u_\ell}) \right|$.
\end{itemize}
The inequality in $w_\ell$ is by the triangle inequality and the definition of $\AMat$: the $\ell$th column $\AMat_\ell$ encodes the descendant leaves of the node $u_\ell$, i.e., $A_{j\ell} = 1_{\{j\in descendant(u_\ell)\cup \{u_\ell\}\}}$. By construction of $\tildebetaTrueVec$, all coefficients in the branch rooted by an aggregating node $u_\ell$ share the same value, i.e., $\beta^*_j = \tilde\beta^*_\ell$ for all $j\in \mathcal L(\mathcal T_{u_\ell})$ where $\ell \in B^*$. Thus, for $\ell \in B^*$, 
$$
\tilde\beta^*_\ell = \frac{\left\| \betaTrueVec_{\mathcal L(\mathcal T_{u_\ell})} \right\|_1}{\left| \mathcal L(\mathcal T_{u_\ell})\right|}
\quad\Rightarrow \quad
\sum_{\ell \in B^*} w_\ell \left|\tilde\beta^*_\ell \right| \le \sum_{\ell \in B^*} \left|\mathcal L(\mathcal T_{u_\ell}) \right|\left( \frac{\left\| \betaTrueVec_{\mathcal L(\mathcal T_{u_\ell})} \right\|_1}{\left| \mathcal L(\mathcal T_{u_\ell})\right|}\right) = \left\|\betaTrueVec \right\|_1.
$$
Plugging the choice of $\lambda$ and the above upper bounds into Theorem \eqref{maintheorem_alpha2} yields
\begin{align*}
\frac{1}{n}\left\| \XMat\betaOurVec - \XMat\betaTrueVec  \right\|_2^2 &\le 12\sigma\sqrt{\frac{\log (2p)}{n}} \left( (1-\alpha)\sum_{j\in\mathcal A^*} \tilde w_j |\beta^*_j| + \alpha\sum_{\ell\in B^*}w_\ell |\tilde\beta^*_\ell|  \right)\\
&\le 12\sigma\sqrt{\frac{\log (2p)}{n}}\left((1-\alpha)\left\|\betaTrueVec \right\|_1 + \alpha\left\|\betaTrueVec \right\|_1 \right)\\
&\lesssim \sigma\sqrt{\frac{\log p}{n}}\left\|\betaTrueVec \right\|_1 \quad (\text{by}\ \log(2p) = \log 2 + \log p\le 2\log p)
\end{align*}
which holds up to a multiplicative constant with probablity at least $1-2/p$.


\section{Proof of Corollary \ref{cor_better_lasso_rate}}
\label{sec:proof_of_cor_better_lasso_rate}

With $\beta^*_1 = \ldots = \beta^*_p$ the aggregating set consists of the root, i.e., $B^* = \{r\}$. Thus, $\tilde\beta^*_r = \beta^*_1 = \|\betaTrueVec\|_1/p$ and $w_r = \|\XMat \bm 1_p\|_2/\sqrt{n}$. With $\|\XMat_j\|_2\le \sqrt{n}$, we have $\tilde w_j = \|\XMat_j\|_2/\sqrt{n}\le 1$ for $1\le j \le p$.

Plugging $\lambda = 4\sigma \sqrt{\log(2p)/n}$ into Theorem \eqref{maintheorem_alpha2} yields
\begin{align*}
\frac{1}{n}\left\| \XMat\betaOurVec - \XMat\betaTrueVec  \right\|_2^2 &\le 12\sigma\sqrt{\frac{\log (2p)}{n}} \left( (1-\alpha)\sum_{j\in\mathcal A^*} \tilde w_j |\beta^*_j| + \alpha w_r |\tilde\beta^*_r|  \right)\\
&\le 12\sigma\sqrt{\frac{\log (2p)}{n}}\left((1-\alpha)\left\|\betaTrueVec \right\|_1 + \alpha\frac{\|\XMat \bm 1_p\|_1}{p\sqrt{n}}\left\|\betaTrueVec \right\|_1 \right)\\
&\lesssim \sigma\sqrt{\frac{\log p}{n}}\left\|\betaTrueVec \right\|_1 \left( (1-\alpha) + \alpha \frac{\left\|\XMat \bm 1_{p} \right\|_2}{p\sqrt{n}} \right)
\end{align*}
which holds up to a multiplicative constant with probablity at least
$1-2/p$. Now, if $\alpha\ge\left[1+\|\XMat \bm
  1_{p}\|_2/(p\sqrt{n})\right]^{-1}$, then
$$
1-\alpha\le \alpha \frac{\left\|\XMat \bm 1_{p} \right\|_2}{p\sqrt{n}} 
$$
and so
$$
\frac{1}{n}\left\| \XMat\betaOurVec - \XMat\betaTrueVec  \right\|_2^2 \lesssim \sigma\sqrt{\frac{\log p}{n}}\left\|\betaTrueVec \right\|_1 \alpha \frac{\left\|\XMat \bm 1_{p} \right\|_2}{p\sqrt{n}}
$$
It follows that if $\|\XMat \bm 1_p\|_2= o(p\sqrt{n})$, then our
method achieves a better rate than the lasso rate of $\sigma\sqrt{\log p /n}\left\|\betaTrueVec \right\|_1 $.


\section{Proof of Proposition \ref{lemma_generalX}}
\label{sec:proof_of_lemma_generalX}

Let $\acute\XMat^{(1)}, \ldots, \acute\XMat^{(p)}\in\{0, 1\}^n$ be \textit{iid} copies of the random vector $\acute \XMat\in\{0, 1\}^n$ with $\bm 1_n^T \acute \XMat=r$ and $\mathbb P(\acute X_i = 1) = r/n$ $\forall i = 1, \ldots, n$. By construction $\{\acute\XMat^{(1)}, \ldots, \acute\XMat^{(p)}\}$ each has $r$ nonzero entries. For the $j$th column of $\XMat$, let $\XMat_j = \sqrt{n/r} \acute \XMat^{(j)}$ so that its nonzero entries equal $\sqrt{n/r}$ and $\|\XMat_j\|_2 = \sqrt{n}$.

For the random vector $\acute \XMat$, its expectation and covariance ($\SigmaMat$) diagonal entries are
$$
\mathbb E\acute\XMat = \frac{r}{n}\bm 1_n \quad\text{and}\quad \Sigma_{ii} = Var(\acute X_i) = \frac{r}{n}\left(1-\frac{r}{n}\right) =:\tau^2.
$$
Suppose $\|\acute \XMat - \mathbb E\acute \XMat \|_2\le L$ for some $L>0$. Let $\ZMat := \sum_{j=1}^p \acute\XMat^{(j)}$ and $v(\ZMat) := \max\{ \|\mathbb E[(\ZMat - \mathbb E \ZMat)(\ZMat - \mathbb E \ZMat)^T]\|_{op}, \mathbb E[\|\ZMat - \mathbb E \ZMat\|_2^2] \}$. Corollary 6.1.2 of \cite{tropp2015} gives the concentration inequality for $\ZMat$: $\forall t\ge 0,$
\begin{equation}\label{eq_tropp1}
\mathbb P\left(\|\ZMat - \mathbb E \ZMat\|_2\ge t \right)\le (n+1) \cdot\exp\left( \frac{-t^2/2}{v(\ZMat) + Lt/3} \right).
\end{equation}

First, we derive a closed form for $L$. By construction of $\acute \XMat$ (in particular $\bm 1_n^T\acute\XMat=r$),
$$
\left\|\acute \XMat - \mathbb E \acute\XMat\right\|_2^2
=\left\|\acute \XMat - \frac{r}{n}\bm 1_n \right\|_2^2
= (n - r) (0 - \frac{r}{n})^2 + r(1-\frac{r}{n})^2
= r(1-\frac{r}{n}) = n\tau^2.
$$
So choose $L=\|\acute\XMat - \mathbb E\acute\XMat\|_2 =\sqrt{n}\tau $. Next, we simplify the two components of $v(\ZMat)$:
\begin{align*}
&\|\mathbb E[(\ZMat - \mathbb E \ZMat)(\ZMat - \mathbb E \ZMat)^T]\|_{op}
= \|Cov(\ZMat)\|_{op} = p \|\SigmaMat\|_{op},\\
&\mathbb E[\|\ZMat - \mathbb E \ZMat\|_2^2] = \sum_{i=1}^n Var(Z_i) = tr(Cov(\ZMat)) = p \cdot tr (\SigmaMat).
\end{align*}
Furthermore, $\max\{\|\SigmaMat\|_{op}, tr(\SigmaMat)\} =
tr(\SigmaMat)$ since $\SigmaMat$ is positive semidefinite. Combining the above yields
$$
v(\ZMat) = p \cdot tr(\SigmaMat) = p \cdot n \cdot \tau^2 = pL^2.
$$

Substituting $v(\ZMat)$ and $\mathbb E \ZMat = pr/n\cdot \bm 1_n$ into \eqref{eq_tropp1} yields
$$
\mathbb P\left(\left\|\ZMat - \frac{pr}{n}\bm 1_n  \right\|_2\ge t\right)
\le (n+1)\cdot\exp\left(\frac{-t^2/2}{pL^2 + Lt/3} \right)\quad\forall t\ge 0.
$$
Since $\XMat_j = \sqrt{n/r}\acute \XMat^{(j)}$, we have $\XMat \bm 1_p= \sqrt{n/r}\ZMat$ and $\forall t\ge 0$,
$$
\mathbb P\left(\left\|\XMat \bm 1_p - p\sqrt{\frac{r}{n}}\bm 1_n \right\|_2\ge\sqrt{\frac{n}{r}}t \right)
=\mathbb P\left(\left\|\ZMat - \frac{pr}{n}\bm 1_n  \right\|_2\ge t\right)
\le (n+1)\cdot\exp\left(\frac{-(t/L)^2/2}{p + (t/L)/3} \right)
=:F(t).
$$
Choosing $t=pr/\sqrt{n}$ and by the triangle inequality,
$$
\|\XMat \bm 1_p\|_2 \le \|\XMat \bm 1_p  - p\sqrt{r/n}\cdot \bm 1_n\|_2 + \|p\sqrt{r/n}\cdot \bm 1_n\|_2\le \sqrt{n/r}\cdot t + p\sqrt{r} = 2p\sqrt{r}
$$
holds with probability at least $1- F(pr/\sqrt{n})$. Also $t/L
=pr/(n\tau)$. We next simplify $F(pr/\sqrt{n})$.
\begin{align*}
F\left(\frac{pr}{\sqrt{n}} \right) 
&= (n+1)\cdot \exp\left(\frac{-\frac{1}{2}\left( \frac{pr}{n\tau}\right)^2 }{p + \frac{1}{3} \left( \frac{pr}{n\tau}\right)} \right)
=(n+1)\cdot \exp\left(\frac{-\frac{pr^2}{2n^2\tau^2} }{1 + \frac{r}{3n\tau}} \right)\\
\Rightarrow\quad\log F\left(\frac{pr}{\sqrt{n}} \right) 
&=\log (n+1) - \frac{1}{2}\cdot \frac{p(r/n)^2}{\tau^2 + \frac{1}{3}\cdot\tau\cdot\frac{r}{n}}\\
&\le \log(n+1) -\frac{1}{2}\cdot \frac{p (r/n)^2}{r/n + \frac{1}{3}(r/n)^{3/2}} \quad \left(\text{by }\tau = \sqrt{\frac{r}{n}\left(1-\frac{r}{n}\right) } \le \sqrt{\frac{r}{n}}\right)\\
&\le \log(n+1) - \frac{1}{2}\cdot \frac{p(r/n)}{1+ \frac{1}{3}}\quad \left(\text{by }\frac{r}{n}\le 1 \right)\\
&= \log(n+1) - \frac{3}{8}\cdot\frac{pr}{n}.
\end{align*}
So, $\mathbb P(\|\XMat\bm 1_p\|_2\le 2p\sqrt{r})\ge
1-\exp\left(\log(n+1) - \frac{3pr}{8n}\right)$. Note that
$\frac{r}{n}>\frac{8\log(n+1)}{3p}$ ensures the exponent converges to
negative infinity as $n$ and $p$ increase. Thus, if
$r>\frac{8n\log(n+1)}{3p}$, then
$$
\|\XMat \bm 1_p\|_2 = O_p(p\sqrt{r})\quad \Leftrightarrow \quad \frac{\|\XMat \bm 1_p\|_2}{p\sqrt{n}} = O_p \left(\sqrt{\frac{r}{n}}\right).
$$
So, we have $\|\XMat \bm 1_p\|_2 = o_p(p\sqrt{n})$ if $r= o(n)$ and $r>\frac{8n\log(n+1)}{3p}$.


\section{Paired $t$-tests between our method and {\tt L1-ag-h}}
\label{sec:pairedttests}

To compare the test errors in the first two columns of Table \ref{model_performance} in the main paper, we perform a two-sample $t$-test for each row.  For each of the $n = 40,000$ reviews, a pair of squared prediction errors is calculated, one from each method.  Conditional on the training set, these $n$ pairs can be thought of as independent.  Table \ref{tablepairedttests} shows the resulting p-values.

\begin{table}[h]
\spacingset{1} 
\centering
\begin{tabular}{@{\extracolsep{-2pt}}r|rrrrrrrr}
\hline
prop. &  1\% & 5\% & 10\% & 20\% & 40\% & 60\% & 80\% & 100\%\\
\hline
p-value & $3.95\times 10^{-9}$ & 0.086 & $7.21\times 10^{-7}$ & $7.56\times 10^{-8}$ & $2.2\times 10^{-16}$ & 0.056 & 0.105 & 0.001\\
\hline
\end{tabular}
\caption{Paired $t$-tests between squared prediction errors from our method and {\tt L1-ag-h} on the held-out test set. The prop. values correspond to different training sizes in Table \ref{model_performance}.}
\label{tablepairedttests}
\end{table}


\section{Hierarchical Clustering Tree with ELMo}
\label{sec:elmotree}

Embeddings from Language Models (ELMo) is a state-of-the-art natural language processing framework for representing words with deep contextualized embeddings \citep{peters2018}. ELMo leverages a deep neural network that is pre-trained on a large text corpus and has multiple attractive properties. For our purpose, the most relevant property of ELMo is that its word embeddings account for the {\em context} in which the words appear. In particular, the ELMo embedding assigned to a word is a function of the entire sentence containing that word. This means that the same word in different contexts will have different embeddings. This is a great advantage over traditional word embeddings such as {\em GloVe} \citep{pennington2014}.

Among the training sets listed in Table \ref{model_performance}, we focus on one scenario when $n=33,997$ and $p=5,621$ (which corresponds to 20\% of the entire training reviews). We use TensorFlow \citep{tensorflow2015} and an ELMo model pre-trained on the One Billion World Language Model Benchmark\footnote{\url{http://www.statmt.org/lm-benchmark/}} to get 1,024-dimensional embeddings for all words in the 33,997 training reviews and the 40,000 held-out testing reviews. To establish a fair comparison with existing experimentation, we keep only adjective embeddings in the reviews, which leaves us 952,974 adjective embeddings for training and 1,122,030 adjective embeddings for testing. Again, the same adjective in various sentences has different ELMo embeddings whose proximities encode how the word is used in each context.

We use mini-batch K-Means clustering \citep{sculley2010} to cluster the 952,974 adjectives from training into 6,001 clusters.  The choice of the number of K-Means clusters ensures that the comparison between ELMo and {\it GloVe} (which is based on $p$=5,621) is roughly on the same scale of features. After mini-batch K-Means clustering, we update each cluster centroid to the averaged embeddings for training adjectives assigned to the cluster. We also associate testing adjectives to the clusters by Euclidean distance. After associating adjectives to clusters, we construct a document-cluster matrix for training and testing, respectively.  The $ij$-entry of the document-cluster matrix has the number occurrences of the $j$th cluster in the $i$th review. Each cluster represents a collection of occurrences of adjectives used in a similar context within a review.  We next generate a tree by performing hierarchical clustering on the cluster centroids. The tree and document-cluster matrix is then used in fitting our model along with others such as {\tt L1-ag-h} and {\tt L1-ag-d}.

\bibliography{ref}
\bibliographystyle{apalike}

\end{document}